# Auxetic behavior and acoustic properties of microstructured piezoelectric strain sensors

Maria Laura De Bellis [a,*], and Andrea Bacigalupo [b,*],

[a] *Department of Innovation Engineering, University of Salento, Lecce, Italy*

[b] *IMT School for Advanced Studies Lucca, Italy*

**Abstract**

The use of multifunctional composite materials adopting piezo-electric periodic cellular lattice structures with auxetic elastic behavior is a recent and promising solution in the design of piezoelectric sensors. In the present work, periodic anti-tetrachiral auxetic lattice structures, characterized by different geometries, are taken into account and the mechanical and piezoelectrical response are investigated. The equivalent piezoelectric properties are obtained adopting a first order computational homogenization approach, generalized to the case of electro-mechanical coupling, and various polarization directions are adopted. Two examples of in-plane and out-of-plane strain sensors are proposed as design concepts. Moreover, a piezo-elasto-dynamic dispersion analysis adopting the Floquet-Bloch decomposition is performed. The acoustic behavior of the periodic piezoelectric material with auxetic topology is studied and possible band gaps are detected.

## 1 Introduction

Over the past few years the use of wireless sensors and wearable electronics has dramatically grown. These devices are spreading not only to different engineering fields, but also to objects in everyday use. Electro-chemical batteries, that need to be periodically replaced, are making way for alternative solutions as piezoelectric actuators adopted to supply power to devices of smaller and smaller dimensions, [1–3] from the micro- (MEMS) to the nano-scale (NEMS). By exploiting their intrinsic electro-mechanical coupling it is, thus, possible to capture the mechanical energy directly available in the surrounding environment, for example in the form of random vibrations, and convert it into usable electrical energy.

In order to make the so-called *energy harvesting* process appealing for real life applications it is fundamental to design piezoelectric actuators with tailored properties. A wide range of different materials, geometries and working principles have been proposed [4–6] for the design of piezoelectric actuators, in order to obtain optimized performances in terms of sensitivity, lightness and reduced size.

---
[*] Corresponding author

In this framework, an up-and-coming topic seems to be the adoption of piezoelectric materials with auxetic behavior in order to enhance the sensitivity of piezoelectric devices, by exploiting their counter-intuitive deformation behavior. Auxetic materials, being characterized by negative Poisson's ratio, expand laterally when subjected to stretching and contract laterally when compressed, [7,8]. The main, well-known, advantages observed in auxetic elastic materials, i.e. increase of the shear modulus, fracture toughness, high acoustic damping and indentation resistance, can be exploited adopting a piezoelectric material and contribute to improve the piezoelectric behaviour.

In [9], the author proposed to embed piezoelectric ceramic rods within an auxetic polymer matrix characterized by an engineered microstructure. The idea was to adopt the auxetic material as passive phase able to redirect the external stress acting on the piezocomposite in order to obtain enhanced device sensitivity. In [10], a possible application for cubic elemental metals with negative Poisson's is the design of strain sensors by sandwiching a sheet of piezoelectric polymer between two thick auxetic metal electrodes. The metal electrode has the property of amplifying the effect of an applied in-plane uniaxial strain on the sensor sheet area.

Particularly noteworthy are, also, piezoelectrically active porous composites [11,12] that exhibit advantages over standard bulk piezo-ceramics in terms of reduction of acoustic impedance and increase of piezoelectric sensitivity. It has also been proven that a crucial beneficial effect is obtained if the increase in porosity is coupled with the introduction of an ordered microstructure [13], this has paved the way for the use of architectured materials, such as periodic cellular lattice structures, whose microstructure can be properly designed to obtain tuned global properties [14–16].

A further step forward is to consider multifunctional composite materials adopting piezoelectric periodic cellular lattice structures with auxetic elastic behavior. Examples are lattices with re-entrant honeycomb or chiral and anti-chiral topologies [17–22]. In recent contributions it has emerged that, by tuning their internal microstructure, it is possible to obtain a wide range of different responses both considering in plane and out-of-plane [23–27] behaviors.

In [28] an unusual behavior of wave propagation in chiral lattices with piezoelectrics is detected. In [29,30] it is proven that piezoelectric lattices based on bimorph ribs exhibit much higher sensitivity than that of material comprising the lattice ribs. Moreover, in [31] it is shown that by integrating lightweight honeycomb structures within existing piezoelectric configurations it is possible to increase in power to weight ratio of piezoelectric harvesters with respect to standard bulk materials.

The study of these composite materials cannot disregard the evaluation of the overall homogenized response via various computational or asymptotic homogenization techniques available in literature. Among others, noteworthy are the approaches resorting to Cosserat [32–35] or second order [36–39] continuum models.

In the present work, periodic anti-tetrachiral auxetic lattice [22] structures, characterized by different geometries, are taken into account and the mechanical and piezoelectrical response are first investigated, see Figure 1(a). The equivalent piezoelectric properties are obtained adopting a first order computational homogenization approach, generalized to the case of electro-mechanical coupling, and various polarization directions are adopted.

As examples of application two microstructured strain sensors, with periodic anti-tetrachiral beam-lattice configuration, are investigated. The devices are characterized by either planar or spatial behavior and in both cases a direct comparison between analytical and numerical solution in terms of displacements, strains and stresses of the equivalent piezoelectric material is performed.

Finally, a 2D piezo-elasto-dynamic dispersion analysis adopting the Floquet-Bloch decomposition is performed. We extend to piezoelectric materials the analysis of chiral honeycomb lattices to evaluate the properties of the dispersion functions of waves propagating in different directions and to detect

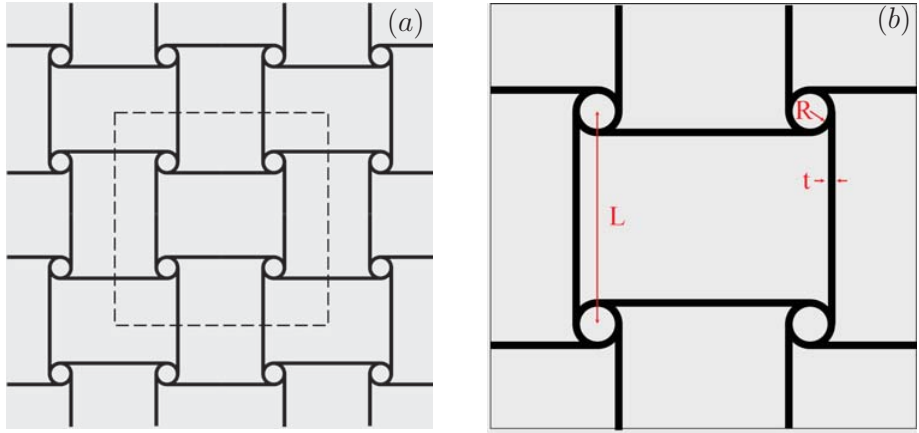

Fig. 1. (a) Schematic of the periodic anti-tetrachiral cellular material with equi-spaced ring connected each by four ligaments. (b) Periodic Cell: L is the distance between the centers of two neighboring rings; R is the radius of the ring and t is the constant thickness of the ligaments.

the band gaps characterizing the material, [22,40–46].

The paper is organized as follows. In Section 2 the geometry of the periodic anti-tetrachiral beam-lattice is briefly described and the governing equations of the piezoelectric material and the mechanical and electrical variables are introduced and commented both at the microscopic scale, Subsection 2.1, and at the macroscopic scale, Subsection 2.2. In Section 2.3, a classical first order computational homogenization technique, generalized to the case of piezoelectric materials, is adopted to derive homogenized constitutive tensors. Section 2.4 is devoted to some illustrative applications. A parametric analysis is first presented and two applications to possible actuators are then shown. In Section 3 the acoustic behavior of the periodic piezoelectric material for different values of the polarization vector **P** is investigated. Finally, in Section 4 some concluding remarks are reported.

## 2 Static analysis of piezoelectric anti-tetrachiral composite material

In Figure 1(b) the geometry of the square periodic cell $\mathcal{A} = [-d/2, d/2] \times [-d/2, d/2]$, with edge $d$, is schematically reported: four rings of radius R are centered at the corners of an ideal square of side L and are interconnected by tangent ligaments. Rings and ligaments are made of piezoelectric material, have the same width t and are possibly filled with a matrix of linear elastic material. In the following, the position vector of a generic microscopic material point $\mathbf{x} = x_1 \mathbf{e}_1 + x_2 \mathbf{e}_2$ is referred to a system of coordinates with origin at point $O$ and orthogonal base ($\mathbf{e}_1, \mathbf{e}_2$). The macroscopic material point is referred to as $\mathbf{X} = X_1 \mathbf{e}_1 + X_2 \mathbf{e}_2$ and the periodic cell is selected such that $\mathbf{X} = \mathbf{0}$ coincides with its geometric center. In order to apply the well established first order computational homogenization approach, the material behavior is described adopting two scales of interest: a macro-scale, the structural one, in which the material is studied as a homogenized medium and a micro-scale where the heterogeneous material is described in detail.

In the following the governing equations together with the boundary conditions at both microscopic and macroscopic scales are reported.

*2.1 Governing equations at the microscopic scale*

The continuum is described as a linear piezoelectric Cauchy medium subject to stresses induced by body forces and free charge densities. For the sake of simplicity, we consider the 2D case. Each material point is characterized by the displacement field $\mathbf{u}(\mathbf{x})$ and the electric potential field $\phi(\mathbf{x})$. The stress tensor $\boldsymbol{\sigma}(\mathbf{x})$ and the electric displacement field $\mathbf{d}(\mathbf{x})$ satisfy the following linear momentum balance and Gauss law, respectively:

$$\begin{aligned} \nabla \cdot \boldsymbol{\sigma}(\mathbf{x}) + \mathbf{b}(\mathbf{x}) &= \mathbf{0}, \\ \nabla \cdot \mathbf{d}(\mathbf{x}) - \rho_e(\mathbf{x}) &= 0, \end{aligned} \tag{1}$$

with $\mathbf{b}(\mathbf{x})$ being the body forces and $\rho_e(\mathbf{x})$ the free charge densities. In particular, the characteristic length associated with the variation of these applied source terms is required to be much greater than the dimension of the periodic cell, i.e. it is required that the principle of separation of scales holds. The coupled constitutive relations in the stress-charge form reads as:

$$\begin{aligned} \boldsymbol{\sigma}(\mathbf{x}) &= \mathbb{C}^m(\mathbf{x})\boldsymbol{\varepsilon}(\mathbf{x}) + \mathbf{e}^m(\mathbf{x})\nabla\phi(\mathbf{x}), \\ \mathbf{d}(\mathbf{x}) &= \widetilde{\mathbf{e}}^m(\mathbf{x})\boldsymbol{\varepsilon}(\mathbf{x}) - \boldsymbol{\beta}^m(\mathbf{x})\nabla\phi(\mathbf{x}), \end{aligned} \tag{2}$$

where $\boldsymbol{\varepsilon}(\mathbf{x}) = sym\nabla\mathbf{u}(\mathbf{x}) = \frac{1}{2}[\nabla\mathbf{u}(\mathbf{x}) + \nabla^T\mathbf{u}(\mathbf{x})]$ is the micro strain tensor, $\mathbf{e}_\phi = -\nabla\phi(\mathbf{x})$ is the electric field, $\mathbb{C}^m(\mathbf{x})$ is the fourth order micro elasticity tensor, $\boldsymbol{\beta}^m(\mathbf{x})$ is the second order dielettric permittivity tensor, $\mathbf{e}^m(\mathbf{x})$ and $\widetilde{\mathbf{e}}^m(\mathbf{x})$ are the third order piezoelectric stress-charce coupling tensors, with the following relation between the components $\widetilde{e}^m_{ijk} = e^m_{jki}$, and the superscript $m$ refers to the microscale.

The resulting partial differential equations governing the piezoelectric problem, in the body domain $\mathcal{B}$, reads

$$\begin{aligned} \nabla \cdot (\mathbb{C}^m(\mathbf{x})sym\nabla\mathbf{u}(\mathbf{x})) + \nabla \cdot (\mathbf{e}^m(\mathbf{x})\nabla\phi(\mathbf{x})) &= \mathbf{0} \quad \text{with} \quad \mathbf{x} \in \mathcal{B}, \\ \nabla \cdot (\widetilde{\mathbf{e}}^m(\mathbf{x})sym\nabla\mathbf{u}(\mathbf{x})) - \nabla \cdot (\boldsymbol{\beta}^m(\mathbf{x})\nabla\phi(\mathbf{x})) &= 0 \quad \text{with} \quad \mathbf{x} \in \mathcal{B}, \end{aligned} \tag{3}$$

with zero source terms. The boundary conditions are of Dirichlet type and Neumann type:

$$\begin{aligned} \mathbf{u}(\mathbf{x}) &= \bar{\mathbf{u}}(\mathbf{x}) & \text{on} & \quad \partial\mathcal{B}_u, \\ \boldsymbol{\sigma}(\mathbf{x})\mathbf{n} &= \bar{\mathbf{t}}(\mathbf{x}) & \text{on} & \quad \partial\mathcal{B}_t, \\ \phi(\mathbf{x}) &= \bar{\phi}(\mathbf{x}) & \text{on} & \quad \partial\mathcal{B}_\phi, \\ \mathbf{d}(\mathbf{x})\mathbf{n} &= \bar{\psi}(\mathbf{x}) & \text{on} & \quad \partial\mathcal{B}_d, \end{aligned} \tag{4}$$

with prescribed values of displacements $\bar{\mathbf{u}}(\mathbf{x})$, surface tractions $\bar{\mathbf{t}}(\mathbf{x})$, electric potential $\bar{\phi}(\mathbf{x})$, free charge density $\bar{\psi}(\mathbf{x})$ and $\mathbf{n}$ is the outer normal. The boundary $\partial\mathcal{B}$ is defined as $\partial\mathcal{B} = \partial\mathcal{B}_u \cup \partial\mathcal{B}_t = \partial\mathcal{B}_\phi \cup \partial\mathcal{B}_d$, $\partial\mathcal{B}_u \cap \partial\mathcal{B}_t = \varnothing$ and $\partial\mathcal{B}_\phi \cap \partial\mathcal{B}_d = \varnothing$.

*2.2 Governing equations at the macroscopic scale*

At the macroscopic scale the governing equations for the 2D problem are likewise derived. At this level, we consider an equivalent homogeneous linear piezoelectric Cauchy medium subject to stresses induced by body forces and free charge densities. Each material point $\mathbf{X}$ is characterized by the displacement field $\mathbf{U}(\mathbf{X})$ and the electric potential field $\Phi(\mathbf{X})$.

The linear momentum balance and the Gauss law, together with the strain-displacement relations and the constitutive relations are formally the same as at the microscopic scale. Therefore, the partial differential equations with zero source terms, in the domain $\mathcal{B}$, are

$$\nabla \cdot \ \mathbb{C}^M(\mathbf{x}) sym \nabla \mathbf{U}(\mathbf{X}) \ + \nabla \cdot (\mathbf{e}^M(\mathbf{X}) \nabla \Phi(\mathbf{X})) = \mathbf{0}, \quad \text{with} \quad \mathbf{X} \in \mathcal{B},$$
$$\nabla \cdot (\widetilde{\mathbf{e}}^M(\mathbf{x}) sym \nabla \mathbf{U}(\mathbf{X})) - \nabla \cdot (\boldsymbol{\beta}^M(\mathbf{X}) \nabla \Phi(\mathbf{X})) = 0 \quad \text{with} \quad \mathbf{X} \in \mathcal{B}, \quad (5)$$

where the superscript $M$ is referred to the macroscopic scale and $\mathbf{E} = sym \nabla \mathbf{U}(\mathbf{X})$ is the strain tensor and $\mathbf{E}_\Phi = \nabla \Phi(\mathbf{X})$ is the electric field. The Dirichlet type and Neumann type boundary conditions are:

$$\begin{aligned}
\mathbf{U}(\mathbf{X}) &= \overline{\mathbf{U}}(\mathbf{X}) & \text{on} \quad \partial \mathcal{B}_u, \\
\mathbf{\Sigma}(\mathbf{X})\mathbf{n} &= \overline{\mathbf{T}}(\mathbf{X}) & \text{on} \quad \partial \mathcal{B}_t, \\
\Phi(\mathbf{X}) &= \overline{\Phi}(\mathbf{X}) & \text{on} \quad \partial \mathcal{B}_\phi, \\
\mathbf{D}(\mathbf{X})\mathbf{n} &= \overline{\Psi}(\mathbf{X}) & \text{on} \quad \partial \mathcal{B}_d,
\end{aligned} \quad (6)$$

with $\overline{\mathbf{U}}(\mathbf{X})$ being prescribed displacements, $\mathbf{\Sigma}(\mathbf{X})$ the macroscopic stress tensor, $\overline{\mathbf{T}}(\mathbf{X})$ the prescribed surface tractions, $\Phi(\mathbf{X})$ and $\overline{\Phi}(\mathbf{X})$ are the macroscopic electric potential and its prescribed counterpart, $\mathbf{D}(\mathbf{X})$ the macroscopic electric displacement and $\overline{\Psi}(\mathbf{X})$ the prescribed macroscopic free charge density.

*2.3 Multi-scale kinematics and macro-homogeneity condition*

A computational homogenization approach generalized to the case of piezoelectric materials is here adopted. In particular, at both scales the first order model is used to describe the physical-mechanical behavior of the material. The multi-field FE analysis is, thus, exploited to evaluate the overall electro-mechanical properties of piezoelectric cellular solids characterized by anti-tetrachiral topology.
The generalized first order multi-scale scheme is defined in the framework of an approach driven by strains and electric-field. The strain tensor $\mathbf{E}(\mathbf{X})$ and the electric field $\mathbf{E}_\Phi(\mathbf{X})$ are, thus, used as input quantity for the periodic cell and a properly defined Boundary Value Problem (BVP) is solved with periodic boundary conditions (PBCs). The displacement and electric potential fields, solution of the BVP at the typical point $\mathbf{x}$ of the periodic cell, can be defined as the superposition of two fields

$$\begin{aligned}
\mathbf{u}(\mathbf{X}, \mathbf{x}) &= \mathbf{u}^*(\mathbf{X}, \mathbf{x}) + \widetilde{\mathbf{u}}(\mathbf{X}, \mathbf{x}), \\
\phi(\mathbf{X}, \mathbf{x}) &= \phi^*(\mathbf{X}, \mathbf{x}) + \widetilde{\phi}(\mathbf{X}, \mathbf{x}),
\end{aligned} \quad (7)$$

where $\mathbf{u}^*(\mathbf{X}, \mathbf{x}) = \mathbf{E}(\mathbf{X})\mathbf{x}$ and $\phi^*(\mathbf{X}, \mathbf{x}) = \mathbf{E}_\Phi(\mathbf{X}) \cdot \mathbf{x}$ are assigned field depending on the macroscopic variables, while $\widetilde{\mathbf{u}}(\mathbf{X}, \mathbf{x})$ and $\widetilde{\phi}(\mathbf{X}, \mathbf{x})$ are periodic perturbation fields arising from the presence of heterogeneities in the material. The boundary displacements and potentials are prescribed between corresponding points $\mathbf{x}^+$ and $\mathbf{x}^-$ belonging to opposite edges of the Periodic Cell, see Figure 14(a), as:

- Vertices of the Periodic Cell:
  $\mathbf{u}_i(\mathbf{X}, \mathbf{x}) = \mathbf{E}(\mathbf{X})\mathbf{x}_i \qquad \phi_i(\mathbf{X}, \mathbf{x}) = \mathbf{E}_\Phi(\mathbf{X}) \cdot \mathbf{x}_i \qquad i = 1, ..., 4,$
- Points on the edges
  $\mathbf{u}^+(\mathbf{X}, \mathbf{x}) - \mathbf{u}^-(\mathbf{X}, \mathbf{x}) = \mathbf{E}(\mathbf{X})\Delta\mathbf{x}, \qquad \phi^+(\mathbf{X}, \mathbf{x}) - \phi^-(\mathbf{X}, \mathbf{x}) = \mathbf{E}_\Phi(\mathbf{X})\Delta\mathbf{x}, \qquad \Delta\mathbf{x} = \mathbf{x}^+ - \mathbf{x}^-$

The macroscopic quantities are defined as the average over the area $A$ of the respective microscopic ones:

$$\begin{aligned}\boldsymbol{\Sigma}(\mathbf{X}) &= \frac{1}{A}\int_A (\boldsymbol{\sigma}(\mathbf{X},\mathbf{x}))d\mathbf{x}, \quad \mathbf{E}(\mathbf{X}) = \frac{1}{A}\int_A (\varepsilon(\mathbf{X},\mathbf{x}))d\mathbf{x} \\ \boldsymbol{D}(\mathbf{X}) &= \frac{1}{A}\int_A (\boldsymbol{d}(\mathbf{X},\mathbf{x}))d\mathbf{x}, \quad \boldsymbol{E}_\Phi(\mathbf{X}) = \frac{1}{A}\int_A (\mathbf{e}_\phi(\mathbf{X},\mathbf{x}))d\mathbf{x}\end{aligned} \quad (8)$$

The microscopic electric enthalpy of the periodic cell reads as

$$\mathcal{H}_m = \frac{1}{2}\int_A [\varepsilon : (\mathbb{C}^m \varepsilon) - \mathbf{e}_\phi \cdot (\boldsymbol{\beta}^m \mathbf{e}_\phi) - \mathbf{e}_\phi \cdot (\widetilde{\mathbf{e}}^m \varepsilon) - \varepsilon : (\mathbf{e}^m \mathbf{e}_\phi)] d\mathbf{x}. \quad (9)$$

The microscopic strain and electric fields can be expressed in terms of the respective macroscopic fields through

$$\begin{aligned}\varepsilon(\mathbf{X},\mathbf{x}) &= \mathbf{B}^1(\mathbf{X},\mathbf{x})\mathbf{E}(\mathbf{X}) + \mathbf{B}^2(\mathbf{X},\mathbf{x})\mathbf{E}_\Phi(\mathbf{X}) \\ \mathbf{e}_\phi(\mathbf{X},\mathbf{x}) &= \widetilde{\mathbf{B}}^2(\mathbf{X},\mathbf{x})\mathbf{E}(\mathbf{X}) + \mathbf{B}^3(\mathbf{X},\mathbf{x})\mathbf{E}_\Phi(\mathbf{X})\end{aligned} \quad (10)$$

where $\mathbf{B}_i(\mathbf{X},\mathbf{x})$ are localization tensors depending on the periodic field obtained by solving the BVP with PBCs in the periodic cell, see e.g. [22]. In particular, $\mathbf{B}^1(\mathbf{X},\mathbf{x})$ is a fourth order tensor, $\mathbf{B}^2(\mathbf{X},\mathbf{x})$ and $\widetilde{\mathbf{B}}^2(\mathbf{X},\mathbf{x})$ are third order tensors and $\mathbf{B}^3(\mathbf{X},\mathbf{x})$ is a second order tensor.
At the macroscopic level the electric enthalpy takes the form

$$\mathcal{H}_M = \frac{1}{2}[\mathbf{E} : (\mathbb{C}^M \mathbf{E}) - \mathbf{E}_\Phi \cdot (\boldsymbol{\beta}^M \mathbf{E}_\Phi) - \mathbf{E}_\Phi \cdot (\widetilde{\mathbf{e}}^M \mathbf{E}) - \mathbf{E} : (\mathbf{e}^M \mathbf{E}_\Phi)]A. \quad (11)$$

By exploiting a generalized macro-homogeneity condition, establishing that $\mathcal{H}_M = \mathcal{H}_m$, the overall electro-mechanical properties of the piezoelectric material are derived and the components of the overall macroscopic elasticity tensor, of the piezoelectric stress-charge coupling tensors and of the dielectric permittivity tensor, respectively, read as

$$\begin{aligned}C^M_{hkrs} &= \frac{1}{A}\int_A (B^1_{ijhk}C^m_{ijpq}B^1_{pqrs} - \widetilde{B}^2_{ihk}\beta^m_{ip}\widetilde{B}^2_{prs} - \widetilde{B}^2_{ihk}\widetilde{e}^m_{ipq}B^1_{pqrs} - B^1_{ijhk}e^m_{ijp}\widetilde{B}^2_{prs})d\mathbf{x}, \\ e^M_{hkr} &= \frac{1}{A}\int_A (B^1_{ijhk}e^m_{ijp}B^3_{pr} + \widetilde{B}^2_{ihk}\widetilde{e}^m_{ipq}B^2_{pqr} + \widetilde{B}^2_{ihk}\beta^m_{ip}B^3_{pr} - B^1_{ijhk}C^m_{ijpq}B^2_{pqr})d\mathbf{x}, \\ \widetilde{e}^M_{hrs} &= \frac{1}{A}\int_A (B^2_{ijh}\widetilde{e}^m_{ijp}\widetilde{B}^2_{prs} + B^3_{ih}e^m_{ipq}B^1_{pqrs} + B^3_{ih}\beta^m_{ip}\widetilde{B}^2_{prs} - B^2_{ijh}C^m_{ijpq}B^1_{pqrs})d\mathbf{x}, \\ \beta^M_{hr} &= \frac{1}{A}\int_A (B^3_{ih}\beta^m_{ip}B^3_{pr} + B^3_{ih}\widetilde{e}^m_{ipq}B^2_{pqr} + B^2_{ijh}e^m_{ijp}B^3_{pr} - B^2_{ijh}C^m_{ijpq}B^2_{pqr})d\mathbf{x},\end{aligned} \quad (12)$$

it is easy to prove that the relation between the components $\widetilde{e}^M_{ijk} = e^M_{jki}$ holds.

*2.4 Illustrative applications*

In this section, some numerical examples are proposed. In subsection 2.4.1, the homogenized electro-mechanical constitutive tensors, characterizing some representative anti-tetrachiral materials, are determined. In this framework, a parametric analysis is performed in order to deduce the influence of both geometrical and physical properties of the microstructure on the overall constitutive properties. Finally, in subsections 2.4.2 and 2.4.3 two examples of strain sensors, characterized by in-plane and out-of-plane behaviors, are presented. The results are critically commented.

*2.4.1 Equivalent piezoelectric properties*

The in-plane equivalent properties of the piezoelectric material are investigated adopting different geometrical parameters of the lattice micro-structure and considering either the case of cellular solid without or with matrix filling the space included within rings and ligaments. Plane strain conditions are assumed.

Referring to the Periodic Cell of the lattice configuration shown in Figure 1(b), we define a reference geometry characterized by the following parameters: the radius of the ring is $R$ =5 mm, the distance between the centers of two neighboring rings is $L$ =25 mm and the constant thickness of the ligaments is $t$ =1.5 mm.

We assume that the lattice structure is made of Lead Zirconate Titanate (PZT-5A) material, polarized along the $\mathbf{e}_1$ direction, i.e. characterized by the polarization vector $\mathbf{P}_1$ whose direction is defined by the polarization unit vector $\mathbf{p}_1 = \mathbf{P}_1/\|\mathbf{P}_1\| = \mathbf{e}_1$, with and without a rubber like matrix filling the space between rings and ligaments.

The non vanishing components of the micro elasticity tensors are: $C^m_{1111}$ = 1.10867 ·$10^{11}$ Pa, $C^m_{2222}$ = 1.2035 ·$10^{11}$ Pa, $C^m_{1122}$ = 7.5090·$10^{10}$ Pa, $C^m_{1212}$=2.5734·$10^{10}$ Pa. The non vanishing components of the stress-charge coupling tensor are: $\widetilde{e}^m_{111}$=15.7835 C/m², $\widetilde{e}^m_{122}$=-5.3512 C/m², $\widetilde{e}^m_{212}$=$\widetilde{e}^m_{221}$=12.2947 C/m². Regarding the PZT-5A material, the non vanishing components of the dielectric permittivity tensor are $\beta^m_{11}/\varepsilon_0$=826.6, $\beta^m_{22}/\varepsilon_0$=919.1, where $\varepsilon_0$=8.854 ·$10^{-12}$ C/(Vm) is the vacuum permittivity. The rubber like material has elastic constants $E$=100 MPa and $\nu$=0.49, [47].

As a first case we consider the anti-tetrachiral beam-lattice material without matrix. The relevant dimensionless components of the overall elasticity tensor $C^M_{ijhk}/C^{PZT}_{ijhk}$ are plotted against $L/R$ in Figure 2(a), with $C^{PZT}_{ijhk}$ being the respective components of the elasticity tensor of the PZT-5A material. In the numerical simulations we assume a constant value for $R$, while $L$ varies. The blue and the red curves correspond to the components $C^M_{1111}/C^{PZT}_{1111}$ and $C^M_{2222}/C^{PZT}_{2222}$, respectively. The differences between the values are ascribable to the polarization effect on the material, in spite of the geometric symmetries of the anti-tetrachiral material. The green curve refers to the $C^M_{1122}/C^{PZT}_{1122}$ component; the negative values are due to the auxetic behavior of the material. Finally, the black curve is the components $C^M_{1212}/C^{PZT}_{1212}$. The curves are characterized by a monotonic trend, as $L/R$ increases, indeed, the material stiffness decreases.

In Figure 2(b) the dimensionless components of the permittivity tensor $\beta^M_{ij}/\beta^{PZT}_{ij}$ are shown. The blue curve and the red curve are the components $\beta^M_{11}/\beta^{PZT}_{11}$ and $\beta^M_{22}\beta^{PZT}_{22}$, respectively. As $L/R$ increases the components of $\boldsymbol{\beta}^M = \varepsilon_0(\mathbf{I} + \boldsymbol{\chi})$ decrease. The components of the electric susceptibility tensor $\boldsymbol{\chi}$ are, therefore, characterized by an analogous behavior, and this means that the material exhibits a lower ability to polarize in response to the electric field.

Figure 2(c) shows the dimensionless components of the stress-charge coupling tensor $e^M_{111}/e^{PZT}_{111}$ (blue curve), $e^M_{221}/e^{PZT}_{221}$ (red curve) and $e^M_{122}/e^{PZT}_{122}$ (green curve) versus $L/R$. Also in this case a monotonic decreasing trend is observed. The coupling components relating both the stress tensor and the electric field and the electric displacement and the strain tensor considerably decrease as $L/R$ increases. In the considered figure, $\beta^{PZT}_{ij}$ and $e^{PZT}_{ijk}$ are the components of the permittivity tensor and of the stress-charge coupling tensor of the PZT-5A material.

At this point we take into account the anti-tetrachiral beam-lattice material in the presence of matrix. The non vanishing dimensionless components of the overall elasticity tensor $C^M_{ijhk}/C^{PZT}_{ijhk}$, of the permittivity tensor $\beta^M_{ij}/\beta^{PZT}_{ij}$ and of the stress-charge coupling tensor $e^M_{ijk}/e^{PZT}_{ijk}$ are shown in Figure 3. The presence of the rubber like material filling the space included within rings and ligaments has the effect of increasing the values of the components of elastic and stress-charge coupling tensors, while components of the permittivity tensor are almost unchanged.

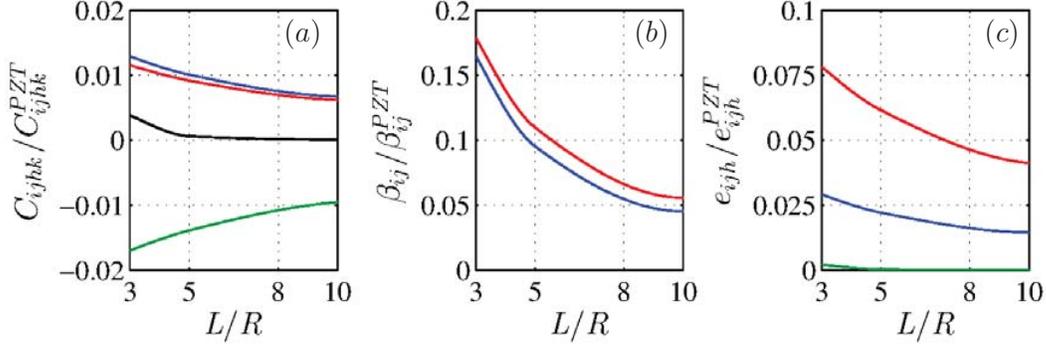

Fig. 2. Piezoelectric equivalent material with polarization direction parallel to the reference unit vector $\mathbf{e}_1$: anti-tetrachiral beam-lattice material without matrix. (a) Components of the equivalent elastic tensor: blue curve $C^M_{1111}$; red curve $C^M_{2222}$; green curve $C^M_{1122}$; black curve $C^M_{1212}$. (b) Components of the equivalent permittivity tensor: blue curve $\beta^M_{11}$; red curve $\beta^M_{22}$. (c) Components of the equivalent stress-charge coupling tensor: blue curve $e^M_{111}$; red curve $e^M_{221}$; green curve $e^M_{122}$.

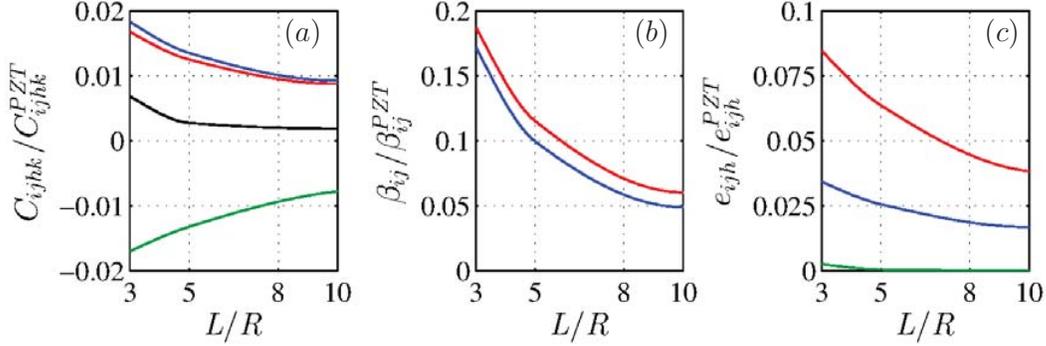

Fig. 3. Piezoelectric equivalent material with polarization vector $\mathbf{P}_1$: anti-tetrachiral beam-lattice material with matrix. (a) Components of the equivalent elastic tensor: blue curve $C^M_{1111}$; red curve $C^M_{2222}$; green curve $C^M_{1122}$; black curve $C^M_{1212}$. (b) Components of the equivalent permittivity tensor: blue curve $\beta^M_{11}$; red curve $\beta^M_{22}$. (c) Components of the equivalent stress-charge coupling tensor: blue curve $e^M_{111}$; red curve $e^M_{221}$; green curve $e^M_{122}$.

We also consider two additional polarization unit vectors $\mathbf{p}_2 = \sqrt{3}/2\mathbf{e}_1 + 1/2\mathbf{e}_2$ and $\mathbf{p}_3 = \sqrt{2}/2\mathbf{e}_1 + \sqrt{2}/2\mathbf{e}_2$ (related to the polarization vectors $\mathbf{P}_2$ and $\mathbf{P}_3$), inclined at angles $\theta_2 = 30°$ and $\theta_3 = 45°$, respectively, with respect to the reference unit vector $\mathbf{e}_1$.

In general, all the components of tensors $\mathbb{C}^M$, $\boldsymbol{\beta}^M$ and $\mathbf{e}^M$ are not vanishing if the polarization vector is not parallel to the unit vectors $\mathbf{e}_1$ and $\mathbf{e}_2$, so that the material symmetries shown for $\mathbf{p}_1$ are modified. The rotation of the polarization vector does not significantly affect the values of the components $C^M_{1111}$, $C^M_{2222}$, $C^M_{1122}$ and $C^M_{1212}$ with respect to the previous case characterized by $\mathbf{p}_1$. The additional dimensionless non vanishing $C^M_{1112}/C^{PZT}_{1112}$ and $C^M_{2212}/C^{PZT}_{2212}$ components are reported in Table 1 for both the polarization unit vectors $\mathbf{p}_2$ and $\mathbf{p}_3$ and the cases of anti-tetrachiral material without and with matrix. As $L/R$ increases, the components in Table 1 monotonically decrease.

Moreover, the components of $\beta^M_{ij}$ and $e^M_{ijk}$ exhibit appreciably different values as the polarization vector $\mathbf{P}$ changes, see Figure 4. In particular, in Figure 4(a) the components of the permittivity tensor $\beta^M_{ij}$ are plotted against $L/R$, in the case of polarization vector $\mathbf{P}_2$ for anti-tetrachiral material without matrix. Here, the component $\beta^M_{12}$, reported in green curve, does not vanish. The values of $\beta^M_{ij}$ are almost the same for both polarization unit vectors $\mathbf{p}_2$ and $\mathbf{p}_3$. In Figure 4(b) the components $e^M_{ijh}$ are shown in the case polarization vector is $\mathbf{P}_2$. In this case all the coupling terms are not negligible. Finally, in the case of $\mathbf{p}_3$, Figure 4(c), as expected, due to the material symmetries, $e^M_{221} = e^M_{112}$,

Table 1
Dimensionless components $C^M_{1112}/C^{PZT}_{1112}$ and $C^M_{2212}/C^{PZT}_{2212}$ versus $L/R$ for polarization unit vectors $\mathbf{p}_2$ and $\mathbf{p}_3$ and for materials without and with matrix.

|  | *without Matrix* | | | | *with Matrix* | | | |
|---|---|---|---|---|---|---|---|---|
|  | $\mathbf{p}_2$ | | $\mathbf{p}_3$ | | $\mathbf{p}_2$ | | $\mathbf{p}_3$ | |
| $\frac{L}{R}$ | $\frac{C_{1112}}{C^{PZT}_{1112}}$ | $\frac{C_{2212}}{C^{PZT}_{2212}}$ | $\frac{C_{1112}}{C^{PZT}_{1112}}$ | $\frac{C_{2212}}{C^{PZT}_{2212}}$ | $\frac{C_{1112}}{C^{PZT}_{1112}}$ | $\frac{C_{2212}}{C^{PZT}_{2212}}$ | $\frac{C_{1112}}{C^{PZT}_{1112}}$ | $\frac{C_{2212}}{C^{PZT}_{2212}}$ |
| 3 | 5.18e−4 | −1.30e−3 | 7.74e−5 | 2.28e−5 | 5.34e−4 | −1.26e−3 | 4.57e−5 | 9.30e−5 |
| 5 | 1.30e−4 | −3.48e−4 | 1.82e−6 | 1.57e−5 | 1.62e−4 | −4.22e−4 | 1.44e−5 | 1.59e−5 |
| 10 | 2.29e−5 | −6.16e−5 | 2.15e−6 | 8.01e−7 | 5.46e−5 | −1.04e−4 | 1.69e−5 | 1.43e−5 |

$e^M_{111} = e^M_{222}$ and $e^M_{121} = e^M_{122}$. In Figure 5, the same plots as in Figure 4 are presented for the case with

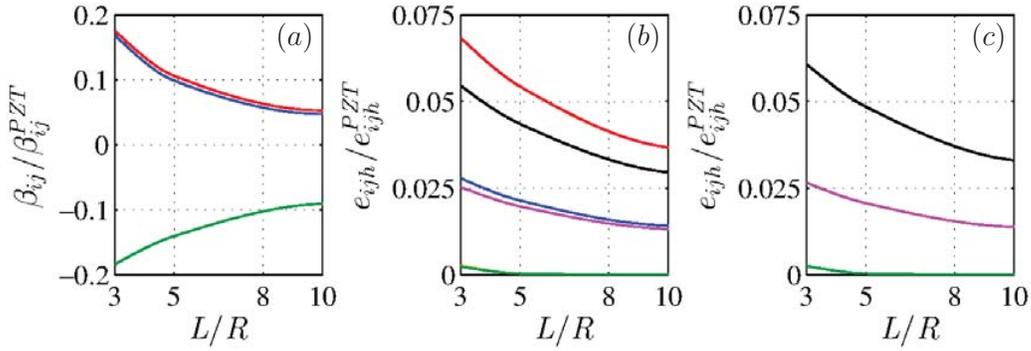

Fig. 4. Piezoelectric equivalent material: lattice microstructure without matrix. Polarization vector $\mathbf{P_2}$ (a) Components of the equivalent permittivity tensor: blue curve $\beta_{11}$; red curve $\beta_{22}$; green curve $\beta_{12}$. (b) Components of the equivalent coupling tensor: blue curve $e_{111}$; red curve $e_{221}$; yellow curve $e_{121}$; black curve $e_{112}$; magenta curve $e_{222}$; green curve $e_{122}$. Polarization vector $\mathbf{P_3}$ (c) Components of the equivalent coupling tensor: blue curve $e_{111}$; red curve $e_{221}$; yellow curve $e_{121}$; black curve $e_{112}$; magenta curve $e_{222}$; green curve $e_{122}$.

matrix. The trends are qualitatively the same as before.

Let $(\mathbf{a}_1, \mathbf{a}_2)$ be the standard basis $(\mathbf{e}_1, \mathbf{e}_2)$ rotated by the counterclock-wise angle $\theta$ about an axis through the origin. An interesting and comprehensive description of the elastic homogenized response can be obtained by evaluating $E^{hom}_\theta$ and $\nu^{hom}_\theta$ related to tension applied only along the direction identified by the unit vector $\mathbf{a}_1$ inclined at an angle $\theta$ with respect to the reference unit vector $\mathbf{e}_1$. The constitutive relations in the strain-charge form result

$$\begin{aligned} \mathbf{E}(\mathbf{X}) &= \mathbb{D}^M(\mathbf{X})\mathbf{\Sigma}(\mathbf{X}) + \mathbf{a}^M(\mathbf{X})\mathbf{D}(\mathbf{X}), \\ -\mathbf{E}_\Phi(\mathbf{X}) &= \widetilde{\mathbf{a}}^M(\mathbf{X})\mathbf{\Sigma}(\mathbf{X}) - \boldsymbol{\alpha}^M(\mathbf{X})\mathbf{D}(\mathbf{X}), \end{aligned} \quad (13)$$

being $\mathbf{E}_\Phi = -\nabla\Phi(\mathbf{x})$, $\mathbb{D}^M(\mathbf{X})$ the elastic compliance tensor, $\boldsymbol{\alpha}^M$ the permittivity tensor at constant stress and $\mathbf{a}^M(\mathbf{X})$ the strain-charge coupling tensor.
In the new rotated frame system, Equations (13) (in component form) become:

$$\begin{aligned} E^\theta_{i_1 j_1} &= D^M_{ijhk} Q_{ii_1} Q_{jj_1} Q_{hh_1} Q_{kk_1} \Sigma^\theta_{h_1 k_1} + a^M_{ijk} Q_{ii_1} Q_{jj_1} Q_{hh_1} D^\theta_{h_1}, \\ -E^{\Phi\text{-}\theta}_{k_1} &= \widetilde{a}^M_{kij} Q_{kk_1} Q_{ii_1} Q_{jj_1} \Sigma^\theta_{i_1 j_1} - \alpha^M_{ki} Q_{kk_1} Q_{ii_1} D^\theta_{i_1} \end{aligned} \quad (14)$$

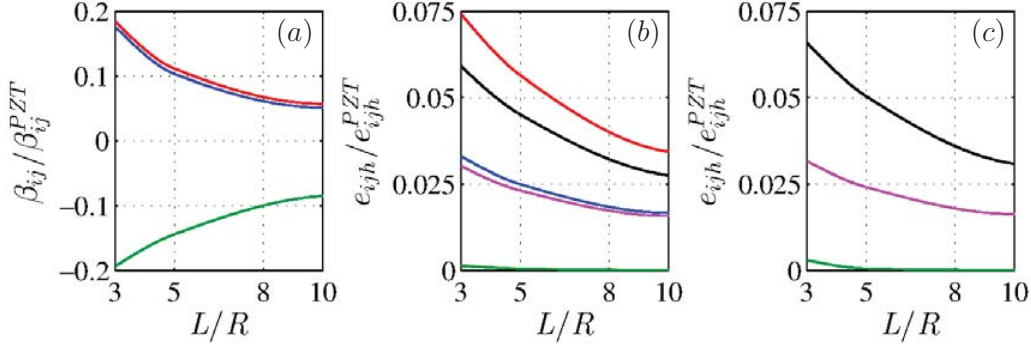

Fig. 5. Piezoelectric equivalent material: lattice microstructure with matrix. Polarization vector $\mathbf{P_2}$ (a) Components of the equivalent permittivity tensor: blue curve $\beta_{11}$; red curve $\beta_{22}$; green curve $\beta_{12}$. (b) Components of the equivalent coupling tensor: blue curve $e_{111}$; red curve $e_{221}$; yellow curve $e_{121}$; black curve $e_{112}$; magenta curve $e_{222}$; green curve $e_{122}$. Polarization vector $\mathbf{P_3}$ (c) Components of the equivalent coupling tensor: blue curve $e_{111}$; red curve $e_{221}$; yellow curve $e_{121}$; black curve $e_{112}$; magenta curve $e_{222}$; green curve $e_{122}$.

where the rotation tensor $\mathbf{Q}$ describes the rotation of $\mathbf{a}_1$ with respect to $\mathbf{e}_1$ and its components are $Q_{11}=Q_{22}=\cos\theta$, $Q_{12}=-Q_{21}=\mathrm{sen}\theta$. In Equations (14) the generic component of constitutive tensors obeys the following transformation law

$$\begin{aligned}
D^{M\_\theta}_{ijhk} &= D^M_{ijhk} Q_{ii_1} Q_{jj_1} Q_{hh_1} Q_{kk_1}, \\
a^{M\_\theta}_{ijk} &= a^M_{ijk} Q_{ii_1} Q_{jj_1} Q_{hh_1}, \\
\widetilde{a}^{M\_\theta}_{ijk} &= \widetilde{a}^M_{ijk} Q_{ii_1} Q_{jj_1} Q_{hh_1}, \\
\alpha^{M\_\theta}_{ki} &= \alpha^M_{ki} Q_{kk_1} Q_{ii_1}
\end{aligned} \tag{15}$$

The overall Young modulus and the Poisson's ratio can be evaluated as a function of $\theta$ as:

$$E^{hom}_\theta = \frac{1}{D^{M\_\theta}_{1111}}, \qquad \nu^{hom}_\theta = -\frac{D^{M\_\theta}_{1122}}{D^{M\_\theta}_{1111}}, \tag{16}$$

with

$$\begin{aligned}
D^{M\_\theta}_{1111} =\, & D^M_{1111}\cos^4\theta + D^M_{2222}\sin^4\theta + 2(D^M_{1122} + 2D^M_{1212})\cos^2\theta\sin^2\theta + \\
& 4D^M_{1211}\cos^3\theta\sin\theta + 4D^M_{2122}\cos\theta\sin^3\theta,
\end{aligned} \tag{17}$$

$$\begin{aligned}
D^{M\_\theta}_{1122} =\, & (D^M_{1111} + D^M_{2222})\cos^2\theta\sin^2\theta + D^M_{1122}\cos^4\theta\sin^4\theta - 4D^M_{1212}\cos^2\theta\sin^2\theta \\
& + 2(D^M_{1211} - D^M_{2122})(\cos\theta\sin^3\theta - \cos^3\theta\sin\theta)
\end{aligned} \tag{18}$$

Analogously, the overall piezoelectric strain-charge coupling constants, related to tension applied only along the direction identified by the unit vector $\mathbf{a}_1$ and zero electric displacement components, as a function of $\theta$ result

$$\widetilde{a}^{hom}_{1\_\theta} = -\widetilde{a}^{M\_\theta}_{111}, \qquad \widetilde{a}^{hom}_{2\_\theta} = -\widetilde{a}^{M\_\theta}_{211}, \tag{19}$$

where $\widetilde{a}^{hom}_{1\_\theta}$ relates $E^{\Phi\_\theta}_1$ to $\Sigma^\theta_{11}$ and $\widetilde{a}^{hom}_{2\_\theta}$ $E^{\Phi\_\theta}_2$ to $\Sigma^\theta_{11}$. That is, they describe the components of the electric field along directions parallel and normal to the direction of application of $\Sigma^\theta_{11}$, respectively.

$$\begin{aligned}
\widetilde{a}^{M\_\theta}_{111} =\, & \widetilde{a}^M_{111}\cos^3\theta + \widetilde{a}^M_{222}\sin^3\theta + (2\widetilde{a}^M_{112} + \widetilde{a}^M_{211})\cos^2\theta\sin\theta + \\
& (2\widetilde{a}^M_{212} + \widetilde{a}^M_{122})\cos\theta\sin^2\theta,
\end{aligned} \tag{20}$$

$$\tilde{a}_{211}^{M,\theta} = -\tilde{a}_{111}^{M}\cos^2\theta\sin\theta + \tilde{a}_{222}^{M}\cos\theta\sin^2\theta - 2\tilde{a}_{112}^{M}\cos\theta\sin^2\theta + \tilde{a}_{211}^{M}\cos^3\theta + 2\tilde{a}_{212}^{M}\cos^2\theta\sin\theta - \tilde{a}_{122}^{M}\sin^3\theta, \quad (21)$$

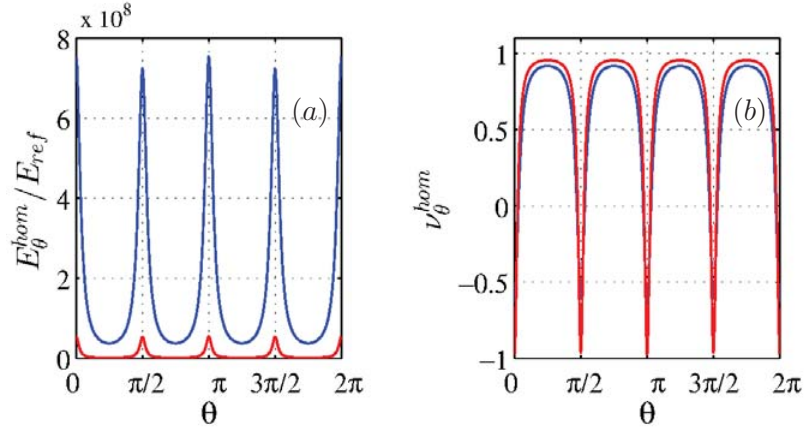

Fig. 6. Equivalent elastic constants versus $\theta$ with polarization vector $\mathbf{P_1}$ and $L/R$=5. (a) $E_\theta^{hom}/E_{ref}$. (b) $\nu_\theta^{hom}$. The red and blue curves are referred to case without and with matrix, respectively.

In Figure 6(a) the elastic modulus $E_\theta^{hom}$, normalized by $E_{ref}$=1 Pa, versus the inclination angle $\theta$ in the case of polarization vector $\mathbf{P_1}$ is shown for the material with matrix (blue line) and without matrix (red line), considering the geometry characterized by $L/R$=5. Similarly, $\nu_\theta^{hom}$ is reported in blue line (with matrix) and in red line (without matrix) in Figure 6(b). A strongly anisotropic behavior is observed with pronounced variations in the elastic properties as $\theta$ slightly varies. It stands to reason that this anti-tetrachiral material, as already highlighted in [22], shows both maximum stiffness and auxeticity along the orthotropy axis. Conversely, when $\theta=n\pi/4$, $n \in \mathbb{Z}$ minimum values of stiffness and auxeticity characterize the material. The presence of the matrix does not modify the material symmetries, but plays a crucial role in increasing the elastic stiffness and in reducing the auxeticity of the material. Concerning the auxeticity, indeed, the Poisson's ratio is equal to -0.97 for the material without matrix and -0.63 for the material with matrix when $\theta=n\pi/2$, $n \in \mathbb{Z}$. Moreover, it is remarkable that the material exhibits auxeticity only for narrow ranges of $\theta$.

In Figure 7, the effect of the polarization direction on $E_\theta^{hom}$ (normalized by $E_{ref}$=1 Pa) and $\nu_\theta^{hom}$ in the case without matrix and $L/R$=5 is investigated. The variation of the polarization direction from $\mathbf{P_1}$ (blue curve) to $\mathbf{P_3}$ (red curve) does not modify significantly the elastic response, a slightly different behavior is detected for the Poisson ratio, even so, outside the range in which the material exhibits auxetic response.

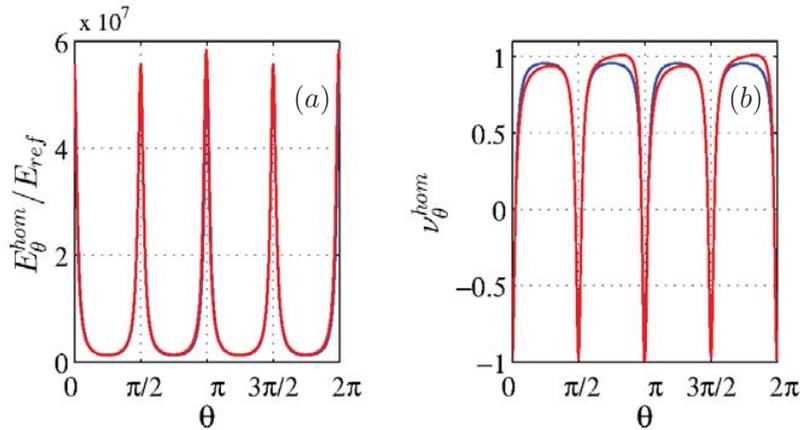

Fig. 7. Equivalent elastic constants versus $\theta$. (a) $E_\theta^{hom}/E_{ref}$. (b) $\nu_\theta^{hom}$. The red and blue curves are referred to case with polarization vector $\mathbf{P_1}$ and $\mathbf{P_3}$, respectively.

In Figure 8(a) the values of $\widetilde{a}_{1\_\theta}^{hom}$ (normalized by $\widetilde{a}_{ref}$=1 $m^2$/C) are plotted versus $\theta$, that is as the direction of the applied stress component varies. The blue curve represents the behavior of the cellular anti-tetrachiral material with matrix, while the red curve is referred to material without matrix. Analogously, Figure 8(b) the values of $\widetilde{a}_{2\_\theta}^{hom}$ (normalized by $\widetilde{a}_{ref}$=1 $m^2$/C) as $\theta$ varies. In both the plots $L/R$=5 and polarirazion vector $\mathbf{P_1}$ are considered. In particular, for a given electric field generated only by a unit stress $\Sigma_{11}^\theta$, $\widetilde{a}_{1\_\theta}^{hom}$ and $\widetilde{a}_{2\_\theta}^{hom}$ represent the components parallel and normal to the stress direction, respectively.

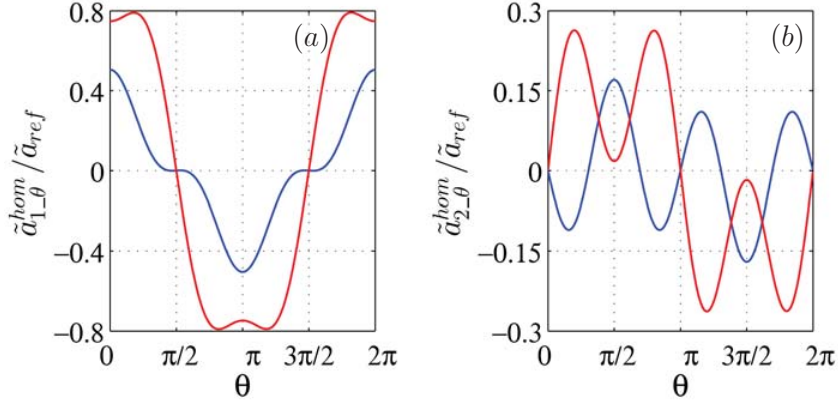

Fig. 8. Equivalent piezoelectric strain-charge coupling constants versus $\theta$, for the case without matrix, with $L/R$=5 and polarirazion vector $\mathbf{P_1}$. (a) $\widetilde{a}_{1\_\theta}^{hom}/\widetilde{a}_{ref}$. (b) $\widetilde{a}_{2\_\theta}^{hom}/\widetilde{a}_{ref}$. The blue and red curves are referred to case with and without matrix, respectively.

### 2.4.2 In-plane auxetic strain sensor

As an example, we consider a strip strain sensor realized by assembling, along directions $\mathbf{e}_1$ and $\mathbf{e}_2$, periodic cells of the anti-tetrachiral PZT-5A piezoelectric material. The strip width is equal to $s$ and the piezoelectric material is characterized by a generic in-plane polarization vector $\mathbf{P}$. Two electrodes are located along the vertical edges of the actuator and a potential difference $\Delta\phi$ is imposed between them. In Figure 9(a) a schematic of the device is shown in the case the material is polarized along the width $s$, i.e. characterized by the direction vector $\mathbf{p}_1$. Besides the micromechanical approach, the overall behavior of the auxetic strain sensor can be also easily described by adopting a first order equivalent homogeneous material, undergoing a potential difference $\Delta\Phi$, as also shown Figure 9(a). In particular, the positive direction of the electric field, generated by the potential difference, corresponds to the direction vector $\mathbf{p}_1$ in the homogenized material.

In the case we consider an infinite strip, the response of the equivalent homogenized material can be described by an analytical solution. Due to the translational invariance in the $\mathbf{e}_2$ direction, the macroscopic displacement and the electric potential, solutions of the field equations, only depend on

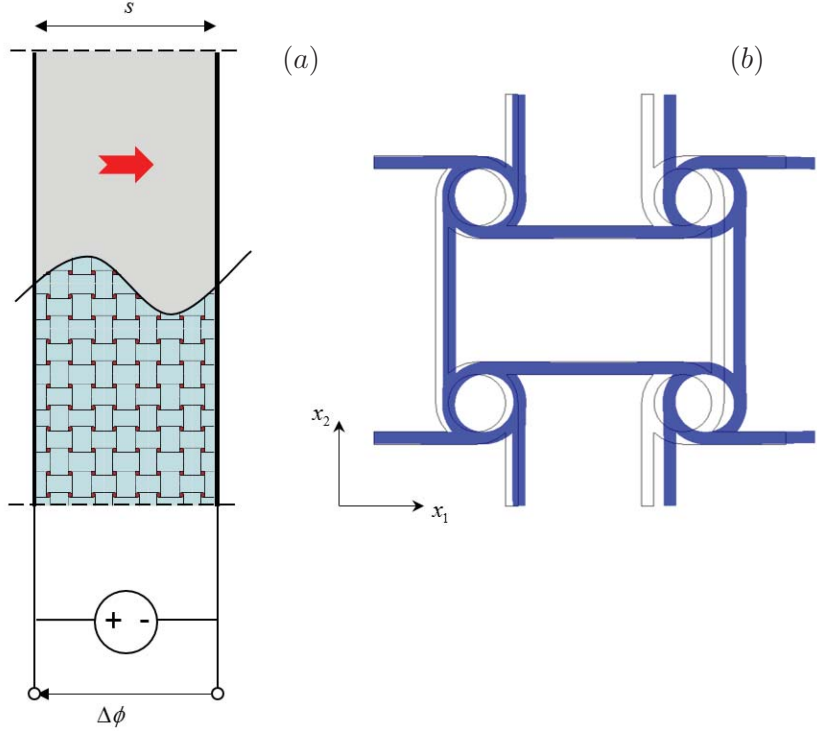

Fig. 9. (a) Schematic of the planar strain sensor. (b) Qualitative deformed shape of a periodic cell when a potential difference $\Delta\phi$ is imposed between the opposite electrodes.

the variable $x_1$. Thus, the field equations, with zero source terms, result as

$$\begin{aligned}
C^M_{1111}U_{1,11} + C^M_{1112}U_{2,11} + e^M_{111}\Phi_{,11} &= 0, \\
C^M_{1112}U_{1,11} + C^M_{1212}U_{2,11} + e^M_{121}\Phi_{,11} &= 0, \\
e^M_{111}U_{1,11} + e^M_{121}U_{2,11} - \beta^M_{11}\Phi_{,11} &= 0,
\end{aligned} \qquad (22)$$

and the boundary conditions are

$$\begin{aligned}
U_1(X_1 = -s/2) &= 0, \\
U_2(X_1 = -s/2) &= 0, \\
\Phi(X_1 = -s/2) &= 0, \\
\Sigma_{11}(X_1 = s/2) &= C^M_{1111}U_{1,1} + C^M_{1112}U_{2,1} + e^M_{111}\Phi_{,1} = 0, \\
\Sigma_{12}(X_1 = s/2) &= C^M_{1112}U_{1,1} + C^M_{1212}U_{2,1} + e^M_{121}\Phi_{,1} = 0, \\
\Phi(X_1 = s/2) &= \Delta\Phi,
\end{aligned} \qquad (23)$$

where $C^M_{ijhk}$ are the components of the elasticity tensor; $e^M_{ijk}$ and $\beta^M_{ij}$ are the components of the piezo-electric stress/charge tensor and the permittivity tensor (at constant strain).

The solution of this ODE problem takes the following form

$$U_1(X_1) = \Delta\Phi \frac{(C_{1112}e_{121} - C_{1212}e_{111})(2X_1 + s)}{2s(C_{1111}C_{1212} + C_{1112}^2)},$$

$$U_2(X_1) = \Delta\Phi \frac{(C_{1111}e_{121} - C_{1112}e_{111})(2X_1 + s)}{2s(C_{1111}C_{1212} + C_{1112}^2)},$$

$$\Phi(X_1) = \Delta\Phi\left(\frac{1}{2} + \frac{X_1}{s}\right) \tag{24}$$

and the $\Sigma_{22}$ stress component is:

$$\Sigma_{22}(X_1) = \Delta\Phi \frac{C_{1111}^M(C_{1212}^M e_{221}^M - C_{2212}^M e_{121}^M) + C_{1122}^M(C_{1112}^M e_{121}^M - C_{1212}^M e_{111}^M)}{s(C_{1111}^M C_{1212}^M - C_{1112}^{M\,2})} +$$

$$+ \Delta\Phi \frac{C_{1112}^M(C_{2212}^M e_{111}^M - C_{1112}^M e_{221}^M)}{s(C_{1111}^M C_{1212}^M - C_{1112}^{M\,2})}. \tag{25}$$

The displacement field **U** in terms of the components of the elastic compliance tensor $\mathbb{D}^M$, the strain-charge coupling tensor $\mathbf{a}^M$ and the permittivity tensor at constant stress $\boldsymbol{\alpha}^M$ read:

$$U_1(X_1) = \Delta\Phi \frac{D_{1122}^M(a_{221}^M \alpha_{22}^M - a_{222}^M \alpha_{12}^M) - D_{2222}^M(a_{111}^M \alpha_{22}^M - a_{112}^M \alpha_{12}^M)\ (2X_1 + s)}{2s\ D_{2222}^M(\alpha_{11}^M \alpha_{22}^M - \alpha_{12}^{M\,2}) + a_{221}^{M\,2}\alpha_{22}^M + a_{222}^{M\,2}\alpha_{11}^M - 2a_{221}^M a_{222}^M \alpha_{12}^M} +$$

$$+ \Delta\Phi \frac{a_{112}^M a_{221}^M a_{222}^M - a_{111}^M a_{222}^{M\,2}\ (2X_1 + s)}{2s\ D_{2222}^M(\alpha_{11}^M \alpha_{22}^M - \alpha_{12}^{M\,2}) + a_{221}^{M\,2}\alpha_{22}^M + a_{222}^{M\,2}\alpha_{11}^M - 2a_{221}^M a_{222}^M \alpha_{12}^M}, \tag{26}$$

$$U_2(X_1) = \Delta\Phi \frac{D_{2212}^M(a_{221}^M \alpha_{22}^M - a_{222}^M \alpha_{12}^M) - D_{2222}^M(a_{121}^M \alpha_{22}^M - a_{122}^M \alpha_{12}^M)\ (2X_1 + s)}{2s\ D_{2222}^M(\alpha_{11}^M \alpha_{22}^M - \alpha_{12}^{M\,2}) + a_{221}^{M\,2}\alpha_{22}^M + a_{222}^{M\,2}\alpha_{11}^M - 2a_{221}^M a_{222}^M \alpha_{12}^M} +$$

$$+ \Delta\Phi \frac{a_{122}^M a_{221}^M a_{222}^M - a_{121}^M a_{222}^{M\,2}\ (2X_1 + s)}{2s\ D_{2222}^M(\alpha_{11}^M \alpha_{22}^M - \alpha_{12}^{M\,2}) + a_{221}^{M\,2}\alpha_{22}^M + a_{222}^{M\,2}\alpha_{11}^M - 2a_{221}^M a_{222}^M \alpha_{12}^M}, \tag{27}$$

and the non-vanishing stress component is:

$$\Sigma_{22}(X_1) = \Delta\Phi \frac{(a_{221}^M \alpha_{22}^M - a_{222}^M \alpha_{12}^M)}{s\ D_{2222}^M(\alpha_{11}^M \alpha_{22}^M - \alpha_{12}^{M\,2}) - 2a_{221}^M a_{222}^M \alpha_{12}^M + a_{222}^{M\,2}\alpha_{11}^M}, \tag{28}$$

In the case the polarization vector is $\mathbf{P_1}$, macroscopic displacement components are

$$U_1(X_1) = \Delta\Phi \frac{(D_{1122}^M a_{221}^M - D_{2222}^M a_{111}^M)(2X_1 + s)}{2s(D_{2222}^M \alpha_{11}^M + a_{221}^{M\,2})},$$

$$U_2(x_1) = 0. \tag{29}$$

and the non vanishing stress component is

$$\Sigma_{22}(X_1) = \Delta\Phi \frac{a_{221}^M}{s(D_{2222}^M \alpha_{11}^M + a_{221}^{M\,2})}. \tag{30}$$

Due to the particular geometry, the Equations (29) coincide with the solution obtained in the case of a strip of finite height whose $U_2$ displacement components are set equal to zero along the horizontal edges.

The planar behavior of the strain sensor is driven by the the sign of $\Delta\Phi$ causing a possible elongation or a shortening along $X_1$, see Figure 9(b) where the deformed shape of a periodic cell is qualitatively depicted.

The auxetic nature of the material has a key role in enhancing the sensitivity of the device for a given $\Delta\Phi$. With reference to Equation (29), it is, indeed, easy to see that for an auxetic material, characterized by negative values of $D^M_{1122}$, the two terms in the numerator $D^M_{1122}a^M_{221}$ and $D^M_{2222}a^M_{111}$ are concordant, thus the resulting horizontal displacement, in a generic point of the strip, has a bigger absolute value than the correspondent displacement exhibited by a non auxetic material where $D^M_{1122}$ has a positive value. We also emphasize that the optimal functioning, i.e. maximum sensitivity, of the device is achieved adopting a material characterized by a polarization vector $\mathbf{P}_1$.

In Figure 10(a) the maximum horizontal displacement $U_1^{max}$, normalized by the corresponding value $U_1^{PZT}$ obtained if the strip were made of bulk PZT-5A material, versus $L/R$ is shown. For all the considered geometries, the equivalent piezoelectric constants are evaluated and the overall homogenized behavior is studied, both for the case with and without matrix between the ligaments, considering a plane strain state. The red and blue curves are referred to the cellular anti-tetrachiral material without matrix and with matrix, respectively. It is noteworthy that the maximum values of the two curves are reached for different geometries ($L/R$=3 for the case without matrix, and $L/R$=5 for the case with matrix). The device realized adopting the anti-tetrachiral microstrusture without matrix turns out to be about 2.25 times more sensitive than the corresponding one made of bulk PZT-5A material for the geometry with $L/R$=3. In the case with matrix, instead, the anti-tetrachiral microstructure proves to be about 1.9 times more sensitive than the device made of bulk PZT-5A material for $L/R$=5. Figure 10(b) shows the maximum vertical stress $\Sigma_{22}^{max}$, normalized by the corresponding value for a device made of bulk PZT-5A strip $\Sigma_{22}^{PZT}$, versus $L/R$. As expected, for both the cases with matrix (blue curve) and without matrix (red curve) the $\Sigma_{22}^{max}$ stress component monotonically decreases as $L/R$ increases. The presence of the matrix has the clear effect of considerably increasing the in-plane stiffness of the devise, as highlighted by the higher values of the considered stress.

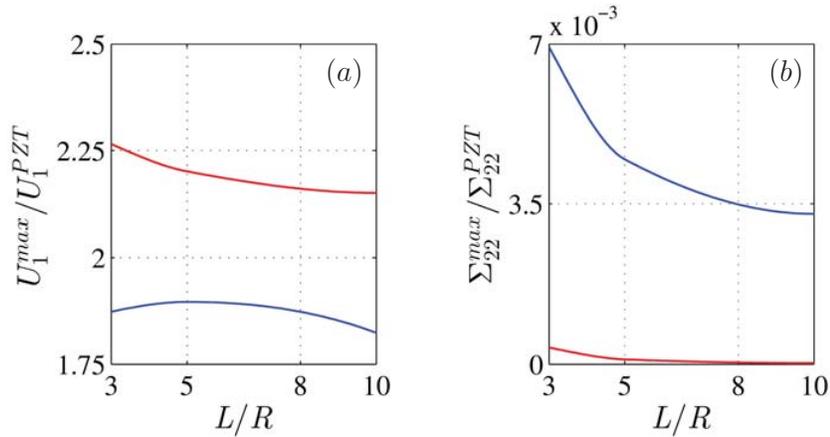

Fig. 10. (a) $U_1^{max}$, normalized by $U_1^{PZT}$ versus $L/R$ for polarization vector $\mathbf{P}_1$. (b) $\Sigma_{22}^{max}$, normalized by $\Sigma_{22}^{PZT}$, versus $L/R$. The blue curves are referred to the material with matrix, while the red curve to the material without matrix.

*2.4.3 Out-of-plane auxetic strain sensor*

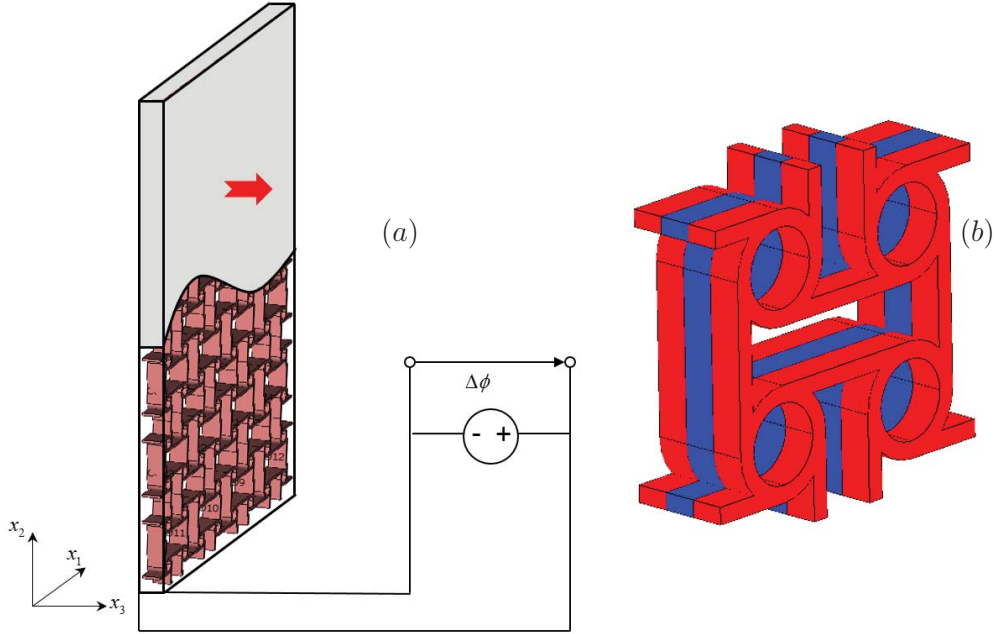

Fig. 11. (a) Schematic of the sandwich structure of the 3D strain sensor; (b) Sketch of a periodic cell with sandwich structure.

As a second example, we consider the strain sensor schematized in Figure 11(a). A generic microscopic material point refers to the position vector $\mathbf{x} = x_1\mathbf{e}_1 + x_2\mathbf{e}_2 + x_3\mathbf{e}_3$ with origin at point $O$ and orthogonal base $(\mathbf{e}_1, \mathbf{e}_2, \mathbf{e}_3)$. The device is made of periodic repetition of cells that exhibit a sandwich structure, as also shown in Figure 11(b): three equal layers (each characterized by thickness s=1/10 $L$) of anti-tetrachiral beam-lattice without matrix are juxtaposed along the thickness in the $\mathbf{e}_3$ direction. The central layer is made of piezoelectric PZT-5A material, while the two outer layers are made of a polymeric material characterized by isotropic decoupled constitutive equations with elastic constants $E$ and $\nu$ and dielectric constant $\beta$. The piezoelectric material is polarized along the $\mathbf{e}_3$ direction, i.e. characterized by the polarization vector $\mathbf{P}_4$ whose direction is defined by the direction vector $\mathbf{p}_4 = \mathbf{P}_4/||\mathbf{P}_4|| = \mathbf{e}_3$. Two electrodes are located on the top and bottom faces of the sandwich structure orthogonal to $\mathbf{e}_3$. The imposition of a difference potential $\Delta\phi$ is responsible for $(\mathbf{e}_1, \mathbf{e}_2)$ in-plane deformations. The positive direction of the electric field, generated by the potential difference, corresponds to the direction vector $\mathbf{p}_4$ in the homogenized material.

The macroscopic behavior of the auxetic strain sensor can be described via a first order equivalent homogeneous material undergoing a difference potential $\Delta\Phi$.

The 3D field equations for the homogenized, with zero source terms, result as

$$\begin{aligned}
&C^M_{1111}U_{1,11} + C^M_{1212}U_{1,22} + C^M_{1313}U_{1,33} + (C^M_{1122} + C^M_{1212})U_{2,12} + \\
&(C^M_{1113} + C^M_{1313})U_{3,13} + (e^M_{113} + e^M_{311})\Phi_{,13} = 0,
\end{aligned}$$

$$\begin{aligned}
&(C^M_{1122} + C^M_{1212})U_{1,12} + C^M_{2222}U_{2,22} + C^M_{1212}U_{2,11} + C^M_{2323}U_{2,33} + \\
&(C^M_{2323} + C^M_{2233})U_{3,23} + (e^M_{223} + e^M_{322})\Phi_{,23} = 0,
\end{aligned}$$

$$\begin{aligned}
&(C^M_{1133} + C^M_{1313})U_{1,13} + (C^M_{2233} + C^M_{2323})U_{2,23} + C^M_{3333}U_{3,33} + C^M_{1313}U_{3,11} + \\
&C^M_{2323}U_{3,22} + e^M_{311}\Phi_{,11} + e^M_{322}\Phi_{,22} + e^M_{333}\Phi_{,33} = 0,
\end{aligned}$$

$$\begin{aligned}
&(e^M_{311} + e^M_{113})U_{1,13} + (e^M_{322} + e^M_{223})U_{1,13} + e^M_{333}U_{3,33} + \\
&e^M_{311}U_{3,11} + e^M_{322}U_{3,22} - \beta_{11}\Phi_{,11} - \beta_{22}\Phi_{,22} - \beta_{33}\Phi_{,33} = 0.
\end{aligned} \qquad (31)$$

We consider two alternative set of boundary conditions modeling different strain sensors characterized either by zero displacement components $U_3$ (SENSOR 1), or zero surface traction $\Sigma_{33}$ (SENSOR 2), on planes $X_3 = \pm s/2$. The boundary conditions for the SENSOR 1 result as

$$\begin{aligned}
&U_1(X_1 = -L/2) = 0, & &\Sigma_{12}(X_1 = -L/2) = 0, & &\Sigma_{13}(X_1 = -L/2) = 0, & &D_1(X_1 = -L/2) = 0, \\
&\Sigma_{11}(X_1 = L/2) = 0, & &\Sigma_{12}(X_1 = L/2) = 0, & &\Sigma_{13}(X_1 = L/2) = 0, & &D_1(X_1 = L/2) = 0, \\
&U_2(X_2 = -L/2) = 0, & &\Sigma_{12}(X_2 = -L/2) = 0, & &\Sigma_{23}(X_2 = -L/2) = 0, & &D_2(X_2 = -L/2) = 0, \\
&\Sigma_{22}(X_2 = L/2) = 0, & &\Sigma_{12}(X_2 = L/2) = 0, & &\Sigma_{23}(X_2 = L/2) = 0, & &D_2(X_2 = L/2) = 0 \\
&U_3(X_3 = -s/2) = 0, & &\Sigma_{13}(X_3 = -s/2) = 0, & &\Sigma_{23}(X_3 = -s/2) = 0, & &\Phi(X_3 = -s/2) = 0, \\
&U_3(X_3 = s/2) = 0, & &\Sigma_{13}(X_3 = s/2) = 0, & &\Sigma_{23}(X_3 = s/2) = 0, & &\Phi(X_3 = -s/2) = \Delta\Phi,
\end{aligned} \qquad (32)$$

The solution of this PDE problem takes the following form

$$\begin{aligned}
U_1(X_1) &= \Delta\Phi \frac{(C^M_{1122}e^M_{223} - C^M_{2222}e^M_{113})(2X_1 + L)}{2s \; C^M_{1111}C^M_{2222} - {C^M_{1122}}^2}, \\
U_2(X_2) &= \Delta\Phi \frac{(C^M_{1122}e^M_{113} - C^M_{1111}e^M_{2233})(2X_2 + L)}{2s \; C^M_{1111}C^M_{2222} - {C^M_{1122}}^2}, \\
U_3(X_3) &= 0, \\
\Phi(X_3) &= \Delta\Phi\left(\frac{1}{2} + \frac{X_3}{s}\right),
\end{aligned} \qquad (33)$$

and the non-vanishing component of stress tensor is

$$\begin{aligned}
\Sigma_{33}(X_3) =& \Delta\Phi \frac{C^M_{1111}(C^M_{2222}e^M_{333} - C^M_{2233}e^M_{223}) + C^M_{1122}(C^M_{1122}e^M_{333} - C^M_{1212}e^M_{111})}{s(C^M_{1111}C^M_{1212} - {C^M_{1112}}^2)} + \\
&+ \Delta\Phi \frac{C^M_{1122}C^M_{2233}e^M_{113} - C^M_{1133}C^M_{2222}e^M_{113}}{s(C^M_{1111}C^M_{1212} - {C^M_{1112}}^2)}.
\end{aligned} \qquad (34)$$

The displacement field $\mathbf{U}$ in terms of the components of the elastic compliance tensor $\mathbb{D}^M$, the strain-charge coupling tensor $\mathbf{a}^M$ and the permittivity tensor at constant stress $\boldsymbol{\alpha}^M$ read:

$$\begin{aligned}
U_1(X_1) &= \Delta\Phi \frac{(D^M_{3333} a_{333} - D^M_{1133} a^M_{333})(2X_1 + L)}{2s(D^M_{3333}\alpha^M_{33} + a^{M\,2}_{333})}, \\
U_2(X_2) &= \Delta\Phi \frac{(D^M_{3333} a_{223} - D^M_{2233} a^M_{333})(2X_2 + L)}{2s(D^M_{3333}\alpha^M_{33} + a^{M\,2}_{333})}, \\
U_3(X_3) &= 0, \\
\Phi(X_3) &= \Delta\Phi\left(\frac{1}{2} + \frac{X_3}{s}\right).
\end{aligned} \qquad (35)$$

and the stress component $\Sigma_{33}$ is

$$\Sigma_{33}(X_3) = -\Delta\Phi \frac{a^M_{333}}{s(D^M_{3333}\alpha^M_{33} + a^{M\,2}_{3333})} \qquad (36)$$

Considering now the SENSOR 2, the boundary conditions are

$$\begin{aligned}
&U_1(X_1 = -L/2) = 0, &&\Sigma_{12}(X_1 = -L/2) = 0, &&\Sigma_{13}(X_1 = -L/2) = 0, &&D_1(X_1 = -L/2) = 0 \\
&\Sigma_{11}(X_1 = L/2) = 0, &&\Sigma_{12}(X_1 = L/2) = 0, &&\Sigma_{13}(X_1 = L/2) = 0, &&D_1(X_1 = L/2) = 0 \\
&U_2(X_2 = -L/2) = 0, &&\Sigma_{12}(X_2 = -L/2) = 0, &&\Sigma_{23}(X_2 = -L/2) = 0, &&D_2(X_2 = -L/2) = 0 \\
&\Sigma_{22}(X_2 = L/2) = 0, &&\Sigma_{12}(X_2 = L/2) = 0, &&\Sigma_{23}(X_2 = L/2) = 0, &&D_2(X_2 = L/2) = 0 \\
&\Sigma_{33}(X_3 = -s/2) = 0, &&\Sigma_{13}(X_3 = -s/2) = 0, &&\Sigma_{23}(X_3 = -s/2) = 0, &&\Phi(X_3 = -s/2) = 0 \\
&\Sigma_{33}(X_3 = s/2) = 0, &&\Sigma_{13}(X_3 = s/2) = 0, &&\Sigma_{23}(X_3 = s/2) = 0, &&\Phi(X_3 = -s/2) = \Delta\Phi
\end{aligned} \qquad (37)$$

in this case the solution of the differential problem is

$$\begin{aligned}
U_1(X_1) =\ & \Delta\Phi \frac{[-C^M_{1122}(C^M_{2233} e^M_{333} - C^M_{3333} e^M_{223}) + C^M_{1133}(C^M_{2222} e^M_{333} - C^M_{2233} e^M_{223})](2X_1 + s)}{2s\ C^M_{1111}(C^M_{2222} C^M_{3333} - C^{M\,2}_{2233}) - C^2_{1122} C^M_{3333} + 2C^M_{1122} C^M_{1133} C^M_{2233} - C^{M\,2}_{1133} C^M_{2222}} + \\
& + \Delta\Phi \frac{e^M_{113}(C^M_{2222} C^M_{3333} - C^{M\,2}_{2233})(2X_1 + s)}{2s\ C^M_{1111}(C^M_{2222} C^M_{3333} - C^{M\,2}_{2233}) - C^2_{1122} C^M_{3333} + 2C^M_{1122} C^M_{1133} C^M_{2233} - C^{M\,2}_{1133} C^M_{2222}} \\
U_2(X_2) =\ & \Delta\Phi \frac{[-C^M_{1122}(C^M_{1133} e^M_{333} - C^M_{3333} e^M_{113}) + C^M_{2233}(C^M_{1111} e^M_{333} - C^M_{1133} e^M_{113})](2X_2 + s)}{2s\ C^M_{1111}(C^M_{2222} C^M_{3333} - C^{M\,2}_{2233}) - C^2_{1122} C^M_{3333} + 2C^M_{1122} C^M_{1133} C^M_{2233} - C^{M\,2}_{1133} C^M_{2222}} + \\
& + \Delta\Phi \frac{e^M_{223}(C^M_{1111} C^M_{3333} - C^{M\,2}_{1133})(2X_2 + s)}{2s\ C^M_{1111}(C^M_{2222} C^M_{3333} - C^{M\,2}_{2233}) - C^2_{1122} C^M_{3333} + 2C^M_{1122} C^M_{1133} C^M_{2233} - C^{M\,2}_{1133} C^M_{2222}} \\
U_3(X_3) =\ & \Delta\Phi \frac{[-C^M_{1111}(C^M_{2222} e^M_{333} - C^M_{2233} e^M_{223}) + C^M_{1122}(C^M_{1122} e^M_{333} - C^M_{1133} e^M_{223})](2X_3 + s)}{2s\ C^M_{1111}(C^M_{2222} C^M_{3333} - C^{M\,2}_{2233}) - C^2_{1122} C^M_{3333} + 2C^M_{1122} C^M_{1133} C^M_{2233} - C^{M\,2}_{1133} C^M_{2222}} + \\
& + \Delta\Phi \frac{e^M_{113}(C^M_{1133} C^M_{2222} - C^M_{1122} C^M_{2233})(2X_3 + s)}{2s\ C^M_{1111}(C^M_{2222} C^M_{3333} - C^{M\,2}_{2233}) - C^2_{1122} C^M_{3333} + 2C^M_{1122} C^M_{1133} C^M_{2233} - C^{M\,2}_{1133} C^M_{2222}} \\
\Phi(X_3) =\ & \Delta\Phi\ \frac{1}{2} + \frac{X_3}{s}\ .
\end{aligned} \qquad (38)$$

The displacement field $\mathbf{U}$ in terms of the components of the elastic compliance tensor $\mathbb{D}^M$, the strain-charge coupling tensor $\mathbf{a}^M$ and the permittivity tensor at constant stress $\boldsymbol{\alpha}^M$ read:

$$\begin{aligned}
U_1(X_1) &= \Delta\Phi \frac{a_{113}^M(2X_1+s)}{2s\alpha_{33}^M}, \\
U_2(X_2) &= \Delta\Phi \frac{a_{223}^M(2X_2+s)}{2s\alpha_{33}^M}, \\
U_3(X_3) &= \Delta\Phi \frac{a_{333}^M(2X_3+s)}{2s\alpha_{33}^M}, \\
\Phi(X_3) &= \Delta\Phi \left(\frac{1}{2} + \frac{X_3}{s}\right).
\end{aligned} \quad (39)$$

The application of the potential difference $\Delta\Phi$ between the top and bottom faces of the sandwich sensor along $\mathbf{e}_3$ is responsible for non-vanishing displacement components $U_1$ and $U_2$ in both cases of SENSOR 1 and SENSOR 2.

As it is evident by inspecting the Equations (35) and (39), the auxeticity of the cellular anti-tetrachiral material does not affect the overall behavior of sensor. The material, indeed, exhibits a purely ($\mathbf{e}_1$, $\mathbf{e}_2$) in-plane auxetic behavior governed by the component $D_{1122}^M$ of the macroscopic elastic compliance tensor.

In the following we present some results associated to two alternative cases characterized by different materials adopted for the outer layers and by different position of the electrodes.

In the first case an all-polymer percolative composite, consisting of conductive polymer particulates (as PANI) in an insulation polymer matrix, [48–52], is considered. The Young modulus is $E$= 535 MPA and the Poisson's coefficient is $\nu$=0.4. This material exhibits very high values of the dielectric constant, directly proportional to the volume fraction of the PANI material. We assume that $\varepsilon_r^{P/PANI} = \beta/\varepsilon_0$ takes values between 1000 and 10000. The electrodes are located on the outer faces of the sandwich structure. In Figure 13(a) the maximum horizontal displacement $U_1^{max}$ (or $U_2^{max}$ for the symmetry of the material), normalized by the corresponding value $U_1^{PZT}$ (or $U_2^{PZT}$) obtained for bulk PZT-5A material characterized by thickness $3s$, versus $\varepsilon_r^{P/PANI}$ is shown for $L/R$=5. The blue curve is referred to SENSOR 1, while the red one to the SENSOR 2. A monotonically increasing trend is shown in both cases as the PANI volume fraction increases. It is worth noting that for SENSOR 1 the adoption of the anti-tetrachiral sandwich structure is extremely advantageous with respect to a bulk PZT-5A strain sensor characterized by the same thickness in the $\mathbf{e}_3$ direction. The maximum value of the normalized displacement component, reached for $\varepsilon_r^{P/PANI}$, is approximately 15. In the case of SENSOR 2, instead, the sandwich strain sensor is advantageous provided that $\varepsilon_r^{P/PANI}$ >1500. In Figure 13(b) a qualitative deformed shape of a heterogeneous periodic cell is reported.

In the second case, the strain sensor presents the outer layers, of insulating polymer material, characterized by $E$=10000 MPa and $\nu$=0.49 (POLYMER A), and by $E$=10000 MPa and $\nu$=0.49 (POLYMER B), [47]. The two electrodes are now located on the top and bottom faces of the PZT-5A layer orthogonal to $\mathbf{e}_3$.

In Figures 13(a) and 13(b) the maximum horizontal displacement $U_1^{max}$ (or $U_2^{max}$ for the symmetry of the material), normalized by the corresponding value $U_1^{PZT}$ (or $U_2^{PZT}$) obtained for bulk PZT-5A material characterized by thickness $s$, versus $L/R$ is shown in the case of SENSOR 1 and SENSOR 2, respectively. The blue curve is referred to POLYMER A, while the red one to POLYMER B. The results show that the SENSOR 2 exhibits a remarkably more sensitive behaviour than SENSOR 1 for all the considered values of $L/R$. In this case the SENSOR 1 is less sensitive than the sensor made of bulk PZT-5A.

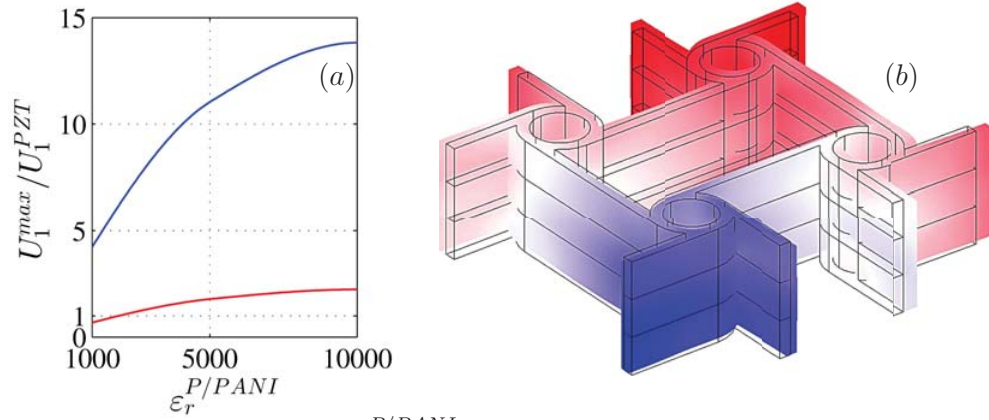

Fig. 12. (a)$U_1^{max}$, normalized by $U_1^{PZT}$ versus $\varepsilon_r^{P/PANI}$. The blue curve is referred to SENSOR 1 and the red curve to SENSOR 2. (b) Deformed shape of a heterogeneous periodic cell.

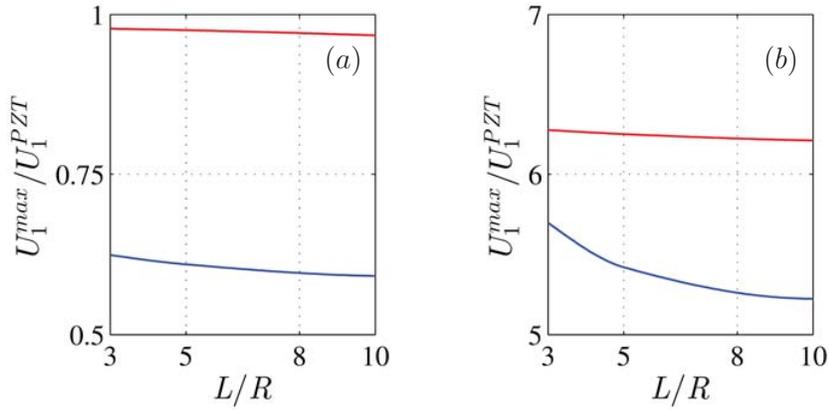

Fig. 13. $U_1^{max}$, normalized by $U_1^{PZT}$ versus $L/R$. The blue curve is referred to POLYMER A and the red curve to POLYMER B. (a) SENSOR 1; (b) SENSOR 2.

## 3  Bulk waves in piezoelectric anti-tetrachiral material

A generalization of the rigorous Floquet-Bloch theory [53] is used to study the dispersion properties of the piezoelectric anti-tetrachiral periodic material. Starting from Equations (1), the set of partial differential equations describing the dynamic balance and the Gauss law of a material point **x** at time $t$, at the microscopic scale, are:

$$\begin{aligned}\nabla \cdot \boldsymbol{\sigma} + \mathbf{b} &= \rho \ddot{\mathbf{u}}, \\ \nabla \cdot \mathbf{d} - \rho_e &= 0,\end{aligned} \quad (40)$$

where $\rho$ is the mass density, $\ddot{\mathbf{u}}$ is the acceleration of the microscopic material point, $\rho_e$ is the free electric charge density and for the sake of brevity the dependence on the **x** has been omitted. By exploiting the constitutive equations (see Equation 2) and the gradient equations $\boldsymbol{\varepsilon} = sym \nabla \mathbf{u} = \frac{1}{2}[\nabla \mathbf{u} + \nabla^T \mathbf{u}]$ and $\mathbf{e}_\phi = -\nabla \phi$, the dynamic governing equations for the piezoelectric material result as

$$\begin{aligned}\nabla \cdot (\mathbb{C}^m sym \nabla \mathbf{u}) + \nabla \cdot (\mathbf{e}^m \nabla \phi) + \mathbf{b} &= \rho \ddot{\mathbf{u}}, \\ \nabla \cdot (\tilde{\mathbf{e}}^m sym \nabla \mathbf{u}) - \nabla \cdot (\boldsymbol{\beta}^m \nabla \phi) - \rho_e &= 0.\end{aligned} \quad (41)$$

By applying the time Fourier transform to (41) in the case of zero source terms, the generalized Christoffel equations are

$$\nabla \cdot (\mathbb{C}^m sym \nabla \widehat{\mathbf{u}}) + \nabla \cdot (\mathbf{e}^m \nabla \widehat{\phi}) + \omega^2 \rho \widehat{\mathbf{u}} = \mathbf{0},$$
$$\nabla \cdot (\widetilde{\mathbf{e}}^m sym \nabla \widehat{\mathbf{u}}) - \nabla \cdot (\boldsymbol{\beta}^m \nabla \widehat{\phi}) = 0. \quad (42)$$

where $\omega$ is the unknown angular frequency and $\widehat{\mathbf{u}}$ and $\widehat{\phi}$ are the time Fourier transform of the displacement field and the electric potential. The generalized Christoffel equations (42) must satisfy proper boundary conditions on the periodic cell. Due to the periodicity of the medium, the following Floquet-Bloch boundary conditions are imposed

$$\widehat{\mathbf{u}}^+ = e^{\mathbf{k}\cdot\Delta\mathbf{x}}\widehat{\mathbf{u}}^-, \quad \widehat{\phi}^+ = e^{\mathbf{k}\cdot\Delta\mathbf{x}}\widehat{\phi}^-, \quad \widehat{\boldsymbol{\sigma}}^+\mathbf{n}^+ = -e^{\mathbf{k}\cdot\Delta\mathbf{x}}\widehat{\boldsymbol{\sigma}}^-\mathbf{n}^-, \quad \widehat{\mathbf{d}}^+\mathbf{n}^+ = -e^{\mathbf{k}\cdot\Delta\mathbf{x}}\widehat{\mathbf{d}}^-\mathbf{n}^- \quad (43)$$

where $\mathbf{k} = k_1\mathbf{e}_1 + k_2\mathbf{e}_2$ is the wave vector, with real components since waves without spatial damping are considered; $\mathbf{n}$ is the outward normal unit vector ($\mathbf{n}^+$ and $\mathbf{n}^-$ are defined on the right/top and on the left/bottom edges, respectively) at $\mathbf{x}^+$ and $\mathbf{x}^-$ on the periodic cell boundary (see Figure 14) and $\Delta\mathbf{x} = \mathbf{x}^+ - \mathbf{x}^-$. Moreover it is possible to define the unit vector of propagation $\mathbf{m}=\mathbf{k}/||\mathbf{k}||$, with $||\mathbf{k}|| = k$ being the wave number. The Floquet–Bloch problem is solved numerically via a Finite Element model. The numerical solution of the differential eigenvalue problem (Christoffel equations (42)) satisfying the Floquet-Bloch boundary conditions in (43) is given in terms of $\omega(\mathbf{k})$, that is the Floquet-Bloch spectrum representing the band structure of the piezoelectric periodic material, and in terms of the eigenfunctions $\widehat{\mathbf{u}}$ and $\widehat{\phi}$ characterizing the polarization of bulk harmonic waves with angular eigenfrequency $\omega$ and wave vector $\mathbf{k}$.

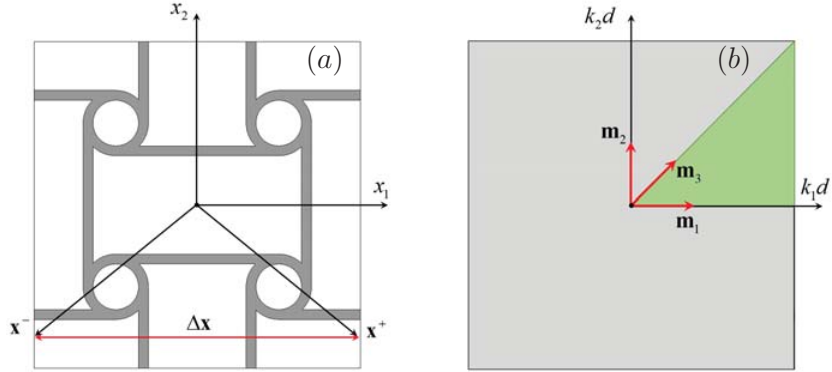

Fig. 14. (a) Periodic boundary conditions: corresponding points $\mathbf{x}^+$ and $\mathbf{x}^-$ on the edges of the periodic cell. The origin of the reference frame is placed in the geometric center of the periodic cell. (b) Brillouin zone (highlighted in light green the first irreducible Brillouin zone) and unit vectors of propagation $\mathbf{m}_i$ (i=1,2,3).

*3.1 Illustrative applications*

As an example, a specific anti-tetrachiral cellular solid without matrix is considered in 2D. The geometry is the same described in Figure 1 with R = 5 mm, t =1.5 mm and L =25 mm. The lattice structure is made of PZT-5A material, whose electro-mechanical properties are reported in Section 2.4.1 and the mass density is $\rho$ = 7750 kg/m$^3$. Three polarization directions are considered, namely the polarization unit vectors are $\mathbf{p}_1 = \mathbf{e}_1$, $\mathbf{p}_3 = \sqrt{2}/2\mathbf{e}_1 + \sqrt{2}/2\mathbf{e}_2$ and $\mathbf{p}_5 = \mathbf{e}_2$, respectively. The influence of both the auxetic properties and the polarization directions on the Floquet–Bloch spectrum, as the unit vector of propagation **m** varies, are here analyzed. Main attention is devoted to identifying the existence, the position and the frequency range of full or partial band gaps in the piezoelectric material. The values of the dimensionless angular frequency $\omega/\omega_{ref}$, with $\omega_{ref}$=1 rad/s, against the dimensionless wave number $k_i d$ (with i=1,2,3), related to the unit vectors of propagation $\mathbf{m}_1 = \mathbf{e}_1$, $\mathbf{m}_2 = \mathbf{e}_2$ and $\mathbf{m}_3 = \sqrt{2}/2 \mathbf{e}_1 + \sqrt{2}/2 \mathbf{e}_2$, are shown in Figures 15, 16 and 17 for polarization unit vectors $\mathbf{p}_1$, $\mathbf{p}_3$ and $\mathbf{p}_5$, respectively. The piezoelectric anti-tetrachiral material exhibits a high spectral density for any considered polarization vectors and the unit vectors of propagation. The first 15 branches of the spectrum are taken into account.

In Figure 15(a) the Floquet-Bloch spectrum for waves characterized by unit vector of propagation

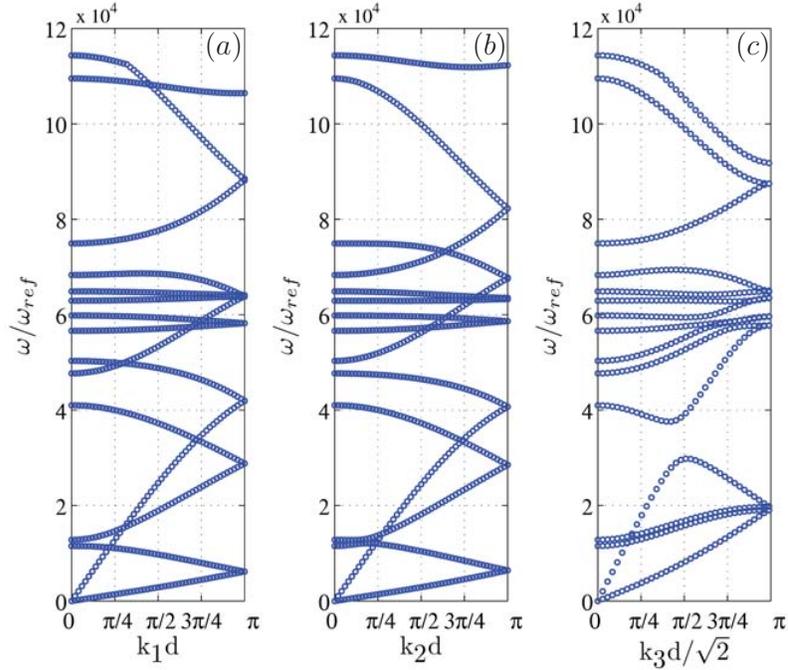

Fig. 15. Floquet-Block spectrum for the anti-tetrachiral piezoelectric material with polarization unit vectors $\mathbf{p}_1 = \mathbf{e}_1$. (a) unit vector of propagation $\mathbf{m}_1$ ; (b) unit vector of propagation $\mathbf{m}_2$ and (c) unit vector of propagation $\mathbf{m}_3$.

$\mathbf{m}_1$ are plotted. Two acoustic branches, departing from $\omega/\omega_{ref}$=0, and the first 13 optical branches are clearly shown. The optical branches present critical points, with zero group velocity, in $k_1 d$=0, i.e. in the long wavelength regime (**k=0**). The Floquet-Bloch spectrum exhibits several points of crossing between acoustic and optical branches and also points of crossing between different optical branches. A partial band gap is detected between the 12th and 13th branches, characterized by a dimensionless amplitude $A_\omega \approx 6630$. In Figure 15(b), instead, the case of unit vector of propagation $\mathbf{m}_2$ is taken into account. In the low frequency range no relevant differences with respect to the case of unit vector of propagation $\mathbf{m}_1$ are observed. Two partial band gaps appear between the 6th and 7th branches, with dimensionless amplitude $A_\omega \approx 2600$, and between the 14th and 15th branches, with dimensionless amplitude $A_\omega \approx 4800$. The relative position of the band gap is shifted towards lower frequencies as for the first band gap, while higher frequencies as for the second band gap with respect to the partial band gap obtained for unit vector of propagation $\mathbf{m}_1$. Moreover, in Figure 15(c) the Floquet-Bloch spectrum for waves characterized by unit vector of propagation $\mathbf{m}_3$ is shown. In this case a lower spectral density is observed in the low frequency range. Between the 2nd acoustic and the 3rd optical branches a veering phenomenon, i.e. a repulsion between the two branches, is observed. This phenomenon engenders a low frequency partial band gap, with dimensionless amplitude $A_\omega \approx 8300$.

A qualitatively similar behaviour has been already detected for the same anti-tetrachiral topology considering a linear elastic non piezoelectric material, see [22]. A second partial band gap, almost coincident with the one observed in Figure 15(a) is also detected.

In Figure 16 the Floquet-Bloch spectra for materials characterized by polarization unit vectors $\mathbf{p}_3$

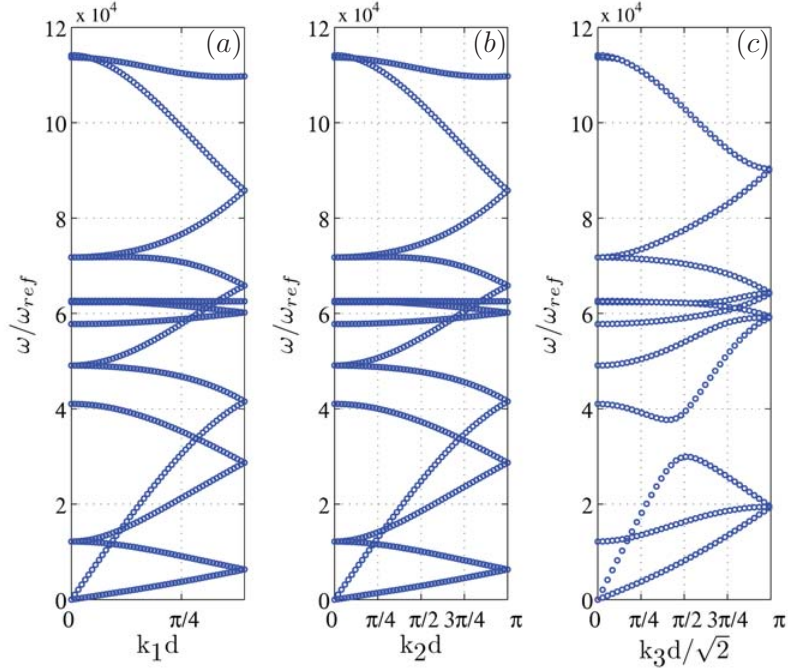

Fig. 16. Floquet-Block spectrum for the anti-tetrachiral piezoelectric material with polarization unit vectors $\mathbf{p}_3 = \sqrt{2}/2\mathbf{e}_1 + \sqrt{2}/2\mathbf{e}_2$. (a) unit vector of propagation $\mathbf{m}_1$; (b) unit vector of propagation $\mathbf{m}_2$ and (c) unit vector of propagation $\mathbf{m}_3$.

are reported. In Figure 16(a) the unit vector of propagation $\mathbf{m}_1$ is considered. Also in this case, two acoustic branches, departing from $\omega/\omega_{ref}=0$, and the first 13 optical branches are shown. The very high spectral density is responsible for the lack of partial band gaps. In Figure 16(b) the unit vector of propagation $\mathbf{m}_2$ is considered. It stands to reason that, by virtue of material symmetry, the same band structure as in Figure 16(a) is detected. The spectrum for waves with unit vector of propagation $\mathbf{m}_3$ is shown in Figure 16(c). Also in this case, between the 2nd acoustic and the 3rd optical branches a veering phenomenon is detected and a low frequency partial band gap, with dimensionless amplitude $A_\omega \approx 8200$, is found.

Finally, in Figure 17 the Floquet-Bloch spectra for materials characterized by polarization unit vectors $\mathbf{p}_5$ are reported. In Figure 17(a) the unit vector of propagation $\mathbf{m}_1$ is considered. A high frequency partial band gap between the 13th and 12th branches, with dimensionless amplitude $A_\omega \approx 3400$, is detected. In Figure 17(b) a unit vector of propagation $\mathbf{m}_2$ is considered. In this case, no relevant band gap appear. Finally, the spectrum for waves with unit vector of propagation $\mathbf{m}_3$ is shown in Figure 17(c). Some band gaps are here detected, four of them are the most relevant and are located between the 2nd acoustic branch and the 3rd optical one (related to a veering phenomena) with $A_\omega \approx 8100$, between the 11th and 12th with $A_\omega \approx 5100$, between the 12th and 13th with $A_\omega \approx 3400$ and between the 13th and 14th with $A_\omega \approx 12300$.

It is noteworthy that the slope of the two acoustic branches, in the long wavelength regime ($\mathbf{k}=\mathbf{0}$), coincide with the phase velocity of shear and compressive waves in a first order homogenized piezoelectric medium. The first order model, indeed, is only able to describe non-dispersive waves characterized by a linear dependence between the angular frequency $\omega$ and the wave-number $k$.

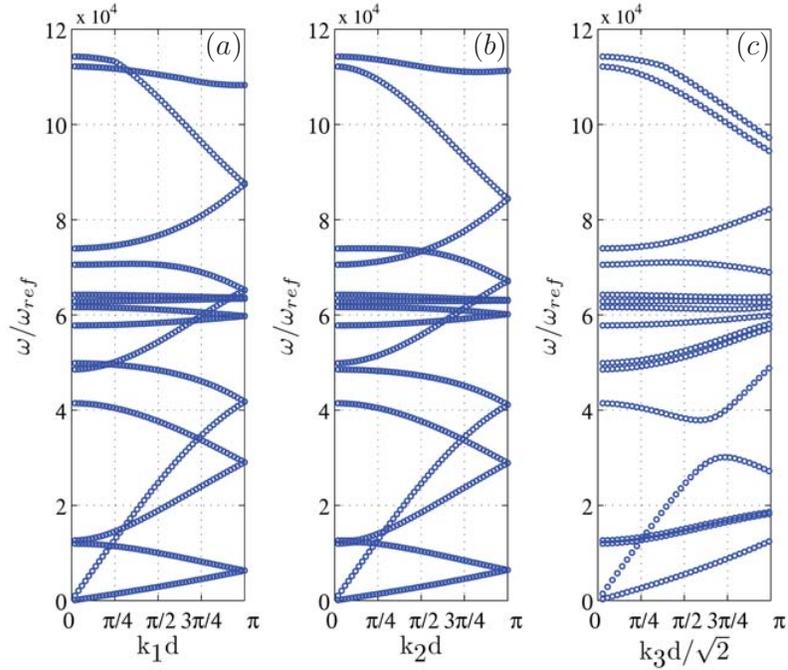

Fig. 17. Floquet-Block spectrum for the anti-tetrachiral piezoelectric material with polarization unit vector $\mathbf{p}_5 = \mathbf{e}_2$. (a) unit vector of propagation $\mathbf{m}_1$; (b) unit vector of propagation $\mathbf{m}_2$ and (c) unit vector of propagation $\mathbf{m}_3$.

In Figure 18 a more comprehensive description of the acoustic characteristics of the piezoelectric anti-tetrachiral material is given by the dispersion surfaces in the Brillouin zone, representing the dimensionless frequency $\omega/\omega_{ref}$ against the dimensionless components of the wave vector ($k_1 d, k_2 d$). In Figure 18(a), 18(b) and 18(c) the polarization unit vectors $\mathbf{p}_1$, $\mathbf{p}_3$ and $\mathbf{p}_5$, respectively, are considered. In the domain of considered wave vectors, a high spectral density is observed since the acoustic surfaces intersect the first optical ones and, moreover, the optical surfaces at higher frequencies intersect each other over and over. It turns out that no total band gaps are detected.

## 4 Final Remarks

Piezoceramic auxetic anti-tetrachiral beam-lattice structures have been investigated. A first order computational homogenization approach has been adopted via a generalization of the macrohomogeneity condition in terms of the electric enthalpy. A parametric analysis has been performed in order to assess the influence of geometric and electro-mechanical parameters on the overall constitutive tensors. In particular, it has been highlighted that a generic polarization vector $\mathbf{P}$ induces an anisotropic material behavior observed both in the elastic and the permittivity tensors. Concerning the auxeticity, analogously to what already found for non piezoelectric materials, it has been detected that the presence of a non piezoelectric matrix within the beam-lattice inhibits the auxetic material behavior. Two strain sensors characterized either by in-plane and out-of-plane behaviour have been analyzed. It has turned out that in the former case (in-plane behavior) the auxetic microstructure makes the strain sensor extremely more sensitive than the case of bulk PZT strain sensor. An enhanced strain capacity may be relevant for sensor and actuator applications. In the latter case (out-of-plane behavior), instead, the auxeticity does not affect at all the increased strain sensitivity of the sensor that is exclusively related to the heterogeneity of the proposed sandwich sensor. In fact, by adopting a 3 layers anti-tetrachiral strain sensor, made of an internal PZT layer and two external ones of an all-polymer percolative composite (consisting of extremely conductive polymer particulates, i.e. PANI, in an insulation polymer matrix), it is possible to obtain very high performances by exploiting the outstanding conductive properties and the high flexibility of the outer layers.

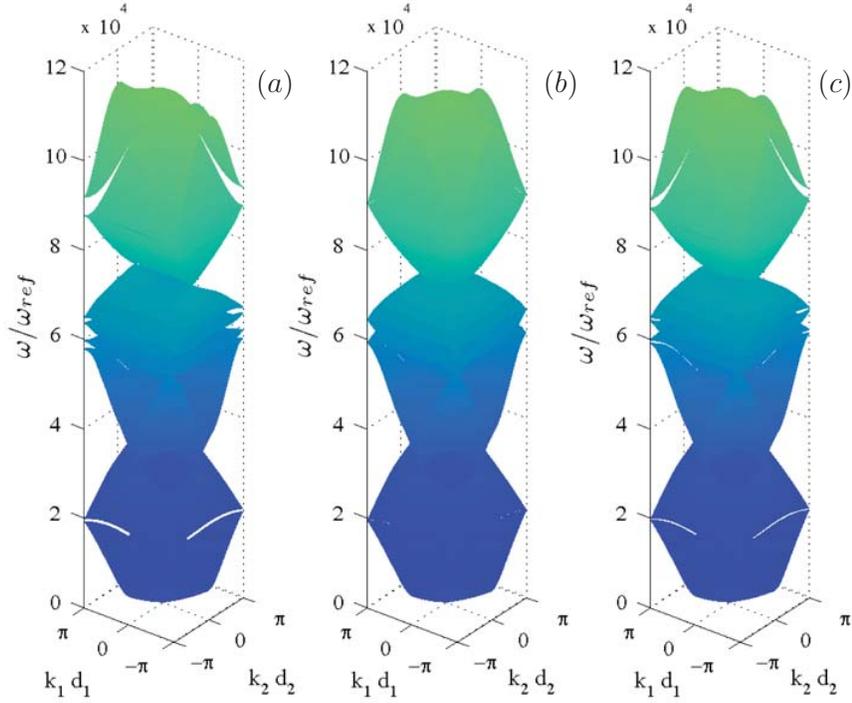

Fig. 18. Dispersive surfaces in the Brillouin zone. (a) polarization unit vector $\mathbf{p}_1$; (b) polarization unit vector $\mathbf{p}_3$ and (c) polarization unit vector $\mathbf{p}_5$.

The acoustic behavior of the periodic piezoelectric material with antitetrachiral topology is studied adopting a generalization of the Floquet-Bloch theory. The analysis of the results in terms of dispersion curves in the Brillouin zone, examined in the dimensionless space, suggest that, the polarization vector $\mathbf{P}$ markedly modify the optical branches at higher frequencies. The piezoelectric anti-tetrachiral material exhibits high spectral densities mainly for waves with unit vector of propagation $\mathbf{m}_1 = \mathbf{e}_1$ and $\mathbf{m}_2 = \mathbf{e}_2$, in fact only high frequency partial band gaps are detected in these cases. For waves with unit vector of propagation $\mathbf{m}_3 = \sqrt{2}/2\mathbf{e}_1 + \sqrt{2}/2\mathbf{e}_2$, instead, the lower spectral density allows the onset of low frequency partial band gaps. Finally, it is worth noting that maximum auxeticity occurs along the $\mathbf{e}_1$ and $\mathbf{e}_2$ axes, while low frequency partial band gaps arise along the direction in which maximum values of the Poisson's coefficients are observed.

**Acknowledgement**

The research leading to these results has received funding from the Regione Puglia under the Future in Research Program "Development of next generation NEMS for energy harvesting"-NSUX1F1. This support is gratefully acknowledged.